\documentclass[transmag]{IEEEtran}
\usepackage{cite}
\usepackage{amsmath,amssymb,amsfonts}
\usepackage{algorithmic}
\usepackage{graphicx}
\usepackage{textcomp}
\usepackage{booktabs}
\usepackage{tabularx}

\usepackage{cite}
\usepackage{amsmath,amssymb,amsfonts}
\usepackage{mathrsfs}
\usepackage{algorithmic}
\usepackage{graphicx}
\usepackage{textcomp}

\usepackage{graphicx}
\usepackage[utf8]{inputenc}
\usepackage[margin=0.75in]{geometry}
\usepackage{mathtools}

\usepackage{mdwmath}
\usepackage{mdwtab}  
\usepackage[ruled,vlined]{algorithm2e}
\SetKwInput{KwInput}{Input}                % Set the Input
\SetKwInput{KwOutput}{Output} 
\newcommand\norm[1]{\left\lVert#1\right\rVert}

\usepackage{cite}
\usepackage{wasysym}
\usepackage{comment}
\DeclareUnicodeCharacter{2061}{}

\DeclareUnicodeCharacter{0301}{}
\usepackage{enumerate}
\DeclareMathOperator*{\argmax}{argmax}
\usepackage{subfigure}
\usepackage{comment}
\usepackage[skip=10pt,font=scriptsize]{caption}

\usepackage{amsfonts}
\usepackage{authblk}
\usepackage{enumitem}
\usepackage{textcomp}

\usepackage{listings}
\usepackage{acronym}
\usepackage{gensymb}
\usepackage{amsmath,amssymb,amsfonts}
\usepackage{algorithmic} 
\usepackage{mathtools}
\usepackage{graphicx} 
\usepackage{mdwmath}
\usepackage{mdwtab}  
\usepackage{textcomp}  
\usepackage[ruled,vlined]{algorithm2e}
\usepackage{enumitem}  
\usepackage[margin=0.75in]{geometry}
\usepackage{wasysym} 
\usepackage{comment}
\usepackage{color,soul}
\DeclareUnicodeCharacter{2061}{}
\DeclareUnicodeCharacter{0301}{}
\DeclareUnicodeCharacter{0015}{}
\usepackage{enumerate}   %\DeclareMathOperator*{\argmax}{argmax}
\usepackage{subfigure}

\usepackage{authblk}
\usepackage{array} \newcolumntype{M}[1]{>{\centering\arraybackslash}m{#1}}

\SetKwInput{KwInput}{Input}                % Set the Input
\SetKwInput{KwOutput}{Output} 

\def\BibTeX{{\rm B\kern-.05em{\sc i\kern-.025em b}\kern-.08em
    T\kern-.1667em\lower.7ex\hbox{E}\kern-.125emX}}

\begin{document}

\title{A Tutorial on Decoding Techniques of Sparse Code Multiple Access }

\author{Saumya Chaturvedi,~Zilong Liu,~Vivek Ashok Bohara,~Anand Srivastava, Pei Xiao.

\thanks{Saumya Chaturvedi,  Vivek Ashok Bohara, and Anand Srivastava are with Indraprastha Institute of Information Technology
(IIIT-Delhi), Delhi, New Delhi, 110020, India (e-mail: \{saumyac, 
vivek.b, anand\}@iiitd.ac.in). Zilong Liu is with the School of Computer Science and Electrical
Engineering, University of Essex, Colchester CO4 3SQ, U.K. (e-mail:
zilong.liu@essex.ac.uk). Pei Xiao is with Institute of Communication Systems, 5G Innovation Centre, University of Surrey, UK (E-mail: {p.xiao@surrey.ac.uk})}\\}
% (\\) to add spacing between authors name and abstract.

%\footnote{}

\IEEEtitleabstractindextext{

\begin{abstract}
Sparse Code Multiple Access (SCMA) is a disruptive  code-domain non-orthogonal multiple access (NOMA) scheme to enable \color{black}future massive  machine-type communication networks. As an evolved variant of code division multiple access (CDMA), multiple users in SCMA are separated by assigning distinctive sparse codebooks (CBs). Efficient multiuser detection is carried out at the receiver by employing the message passing algorithm (MPA) that exploits the sparsity of CBs to achieve error performance approaching to that of the maximum likelihood receiver. In spite of  numerous research efforts  in recent years, a comprehensive one-stop  tutorial of SCMA covering  the background, the basic principles, and new advances, is still missing, to the best of our knowledge. To fill this gap and to stimulate more forthcoming research, we provide a holistic introduction to the principles of SCMA encoding, CB design, and MPA based decoding in a self-contained manner. As an ambitious paper aiming to push the limits of SCMA, we present a survey of advanced decoding techniques with brief algorithmic descriptions as well as several promising directions.  

\end{abstract}


}
\maketitle
\section{INTRODUCTION}
%{\let\thefootnote\relax\footnote{{This work was supported in part by the SPARC grant (SPARC/2018-2019/P950/SL) from Minister of Education Government of India and  by TCS Research Scholarship Program.}}}
\IEEEPARstart{T}{he} next-generation mobile communication  systems are expected to support a wide range of vertical industries such as e-health, smart homes/cities, connected autonomous vehicles, and factories-of-future \cite{uses,scmapot_1,scmapot_2,scmapot_3,scmapot_4}. While this leads to ubiquitous, rapid, and intelligent data services, anytime and anywhere,  the explosive growth of communication devices and applications arising from the 5G-and-beyond  communication networks reinforces the need of a higher spectrum efficiency, reduced access latency, and more reliable data transmission.

Multiple access, as one of the core techniques of wireless communication, aims to enable two or more  users to access  the finite system resources simultaneously in an efficient  manner. Legacy multiuser communication systems mostly rely on  orthogonal multiple access (OMA) schemes where multiple users are mutually orthogonal in a certain domain.  In the past  cellular networks, there have been time division multiple access (TDMA), frequency division multiple access (FDMA), code division multiple access (CDMA), and orthogonal frequency-division multiple access (OFDMA) \cite{scmapot_5}. As an example, in TDMA, each user is given a distinct time slot, and no two users concurrently share  the same time slot. In every OMA scheme, the number of users being simultaneously served is at most  that of the orthogonal resources. Thus, the OMA techniques may not meet the stringent requirements  of next-generation mobile applications.

In recent years, non-orthogonal multiple access (NOMA) has attracted tremendous research attention owing to its capability in supporting more concurrent communication links compared to OMA \cite{scmapot_6,scmapot_7}. %\cite{ad8bbbe54d5f43d88e864886dd63bd15}. 
In a NOMA system, two or more users are superimposed over an identical  physical resource (e.g., power, frequency, time, or code) to provide an overloading factor larger than one. %\cite{Liu2021}.
In general, there are two types of NOMA:  power-domain NOMA (PD-NOMA) and code-domain NOMA (CD-NOMA). In PD-NOMA, multiple users communicating over the same frequency/time resources are separated by assigning them with different power levels \cite{scmapot_8,scmapot_10}. On the other hand, CD-NOMA may be regarded as an extension of CDMA where different users are allocated with distinctive codebooks/sequences. Unlike a traditional CDMA system in which matched filter based receiver is adopted, more advanced receivers may be employed in CD-NOMA.  
%\color{black}

In 2013, sparse code multiple access (SCMA) was proposed by H. Nikopour  and  H. Baligh as a disruptive CD-NOMA scheme \cite{proposed}.  
Building upon the initial idea of low-density signature CDMA (LDS-CDMA) \cite{lds},  several bits of every user are directly mapped to a sparse complex vector (codeword) which is drawn from its associated sparse codebook (CB) \cite{scmapot_11,scmapot_12}. \color{black} Also, SCMA  brings the “constellation  shaping gain” that may not be admissible for LDS-CDMA.
%, where “shaping gain” is the gain in the average symbol energy when the shape of a constellation is changed. 
\color{black}

\subsection{Related Works}
\color{black}Two major research lines are associated with SCMA: \color{black}CB design and multiuser detection (MUD). The goal of CB design is to design and optimize CBs with respect to certain wireless channels against a number of system performance measures such as bit/packet error rate, spectral efficiency, and peak-to-average power ratio (PAPR). After the first CB design work \cite{cb2} in 2014, numerous research attempts have been made. New CBs are designed in \cite{cb3} from \emph{M}-order pulse-amplitude modulation (PAM) to maximize the sum rate of SCMA. In \cite{interleaving},  constellation rotation and interleaving are employed to maximize the minimum Euclidean distance (MED) of sparse CBs. In \cite{starqam}, novel CBs for both Rayleigh and Gaussian channels based on Star-QAM constellations have been developed to minimize the error probability. Inspired by recent  advances in machine learning,  new CBs have been found in \cite{ref4} using the deep-learning-based method by considering minimum bit error rate (BER) as the optimization criteria.  In \cite{gam}, %[17]
golden angle modulation
 constellations have been used to obtain CBs exhibiting excellent error rate performances in  uplink and downlink Rayleigh fading channels. Further, a low complexity construction algorithm for near-optimal CBs has been developed in \cite{nearopt}.
 
Thanks to the sparsity of the CBs, the multiuser signals can be efficiently detected and recovered with the aid of the message passing algorithm (MPA) \cite{lds} whose error rate performance approaches  that of  maximum a posteriori (MAP) detector. In MPA, the belief messages are passed along the edges of the factor graph that characterizes the sparse CBs of an SCMA system. It is shown in \cite{yiqun} that MPA works well even in an SCMA system with an overloading factor of three. In \cite{ref7}, a simpler MPA in the log-domain was introduced by simplifying the exponent and multiplication operations to maximization and addition operations, respectively. In \cite{scmapot_121_35}, a partial marginalization based MPA was proposed for fixed low-complexity SCMA detection. \cite{thresholdmpa} proposed a belief threshold-based MPA in which the belief message updates for every user are checked in every iteration,  thus reducing the total number of computational operations when the number of iterations increases. In \cite{lutmpa}, a lookup table method was designed to simplify the function node update. % In \cite{mimompa}, a hybrid belief propagation and expectation propagation  receiver was proposed for a downlink MIMO-SCMA system. 
In \cite{edgempa}, a threshold-based edge selection and Gaussian approximation (ESGA) was proposed by applying partial Gaussian approximation dynamically for significant complexity reduction at the receiver.
%In \cite{ref7}, max log MPA was proposed which converts multiplication and exponent operations to addition and maximization operation using Jacobian logarithm formula. 
%\underline{\emph{Organization of this paper:}}
\subsection{Motivations and Contributions}

Among many survey/tutorial papers on NOMA in the current literature, most of them focus on PD-NOMA \cite{scmapot_6, scmapot_13_noma_1,scmapot_25_noma_2,scmapot_15_noma_3,scmapot_18_noma_4,scmapot_20_noma_5,scmapot_19_noma_6,scmapot_22_noma_7,scmapot_29_noma_8}.
There is a clear paucity of comprehensive survey papers to SCMA for different levels of readers  \cite{scmapot_14_scma_1,scmapot_28_scma_2,scmapot}.  In \cite{scmapot_14_scma_1}, an overview of  multi-dimensional 
constellations as well as their applications in CB design for uplink SCMA systems  has been  presented.   A book chapter is reported in \cite{scmapot_28_scma_2}, where the basic principles of SCMA are briefly  covered along with CB designing and MUD.  An  introduction to SCMA was given in  \cite{scmapot}, covering an overview survey of   the CB design, detector designs, and some other aspects like resource allocation, practical implementations. However, to the best of our knowledge, a one-stop tutorial-level coverage of SCMA from encoding to decoding is still lacking.

Aiming for influencing both layman beginners and senior scholars and in order to further promote it to a broader community, we present a holistic and systematic introduction to SCMA covering the basic principles of encoding, CB design, and MUD, supplemented with various step-by-step examples, illustration figures,  and mathematical expressions. In particular, we survey a series of advanced SCMA decoding techniques from both MPA and non-MPA aspects and outline a few promising directions for future research on SCMA. We hope that this disruptive CD-NOMA technique can attract more research attention and consequently stimulate new innovative wireless technologies via evolution and/or integration in the era of the 6G research.

%Albeit numerous research attempts on SCMA have been made in the past few years, a comprehensive and in-depth tutorial on the principles of SCMA from a beginner's point of view is missing, to the best of our knowledge. To promote this disruptive multiple access technique and to stimulate more forthcoming research activities on SCMA, we provide a self-contained introduction to SCMA (from CB design to MPA decoding) in a systematic manner. In particular, we explain how efficient MPA decoding is carried out with the aid of a graphical model. 
\subsection{Organization of this work}
The paper's outline is as follows: Firstly, the SCMA system model is presented in Section II, where SCMA encoding and CB designing are discussed. SCMA decoding is discussed  in Section III, which  includes the MAP detection of SCMA, MPA detection, and MPA detection illustrated by an example of a $4\times 6$ SCMA system (i.e., six users communicate concurrently over four resource elements (REs)). Section IV introduces  an array of advanced MPA variants in the literature for both coded and un-coded SCMA systems. Subsequently, we discuss some other non-MPA detection schemes of SCMA, such as expectation propagation and deep learning based SCMA detectors. New directions of SCMA are introduced in Section V.  Finally, this paper is concluded in Section VI. 

\begin{figure*}[htbp]
\centering
\includegraphics[scale=0.4,trim=1cm 1cm 0cm 0.5cm,clip]{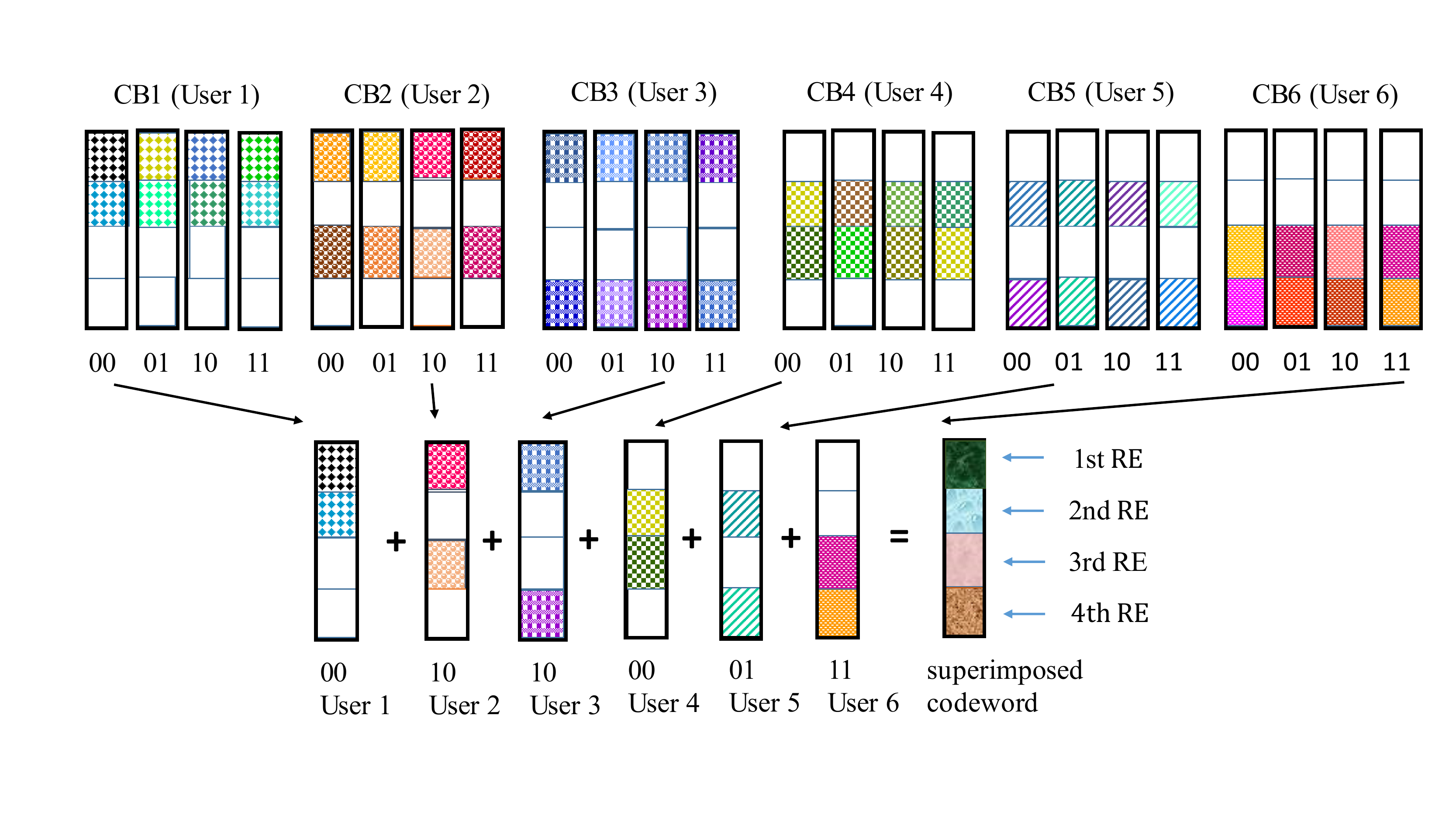}\\
\caption{ An illustration of $4\times 6$ SCMA  Encoding.}
\label{fig:encode}
\end{figure*}

%We start with the basic representation of an SCMA system and CB designing in Section II via the signature matrix. Section III is focused on  decoding using MPA. %In Subsection  3.1, an example on (3,1) repetition code with both hard decoding and soft decoding is discussed showing the advantages of the latter one. 
%Specifically, in Subsections  III-A and III-B, maximum a posteriori (MAP) estimation and factor graphs are introduced, respectively. Subsection III-C presents a high-level introduction on the MPA, whilst subsection III-D demonstrates how belief messages are passed from function nodes to variable nodes and vice-versa in a factor graph. Subsection III-E presents the sum-product algorithm (SPA, a realization approach of MPA) carried out by passing belief messages over a factor graph with the aid of an example. Subsection III-F discusses the SPA and max-log-SPA used for multi-user detection of SCMA signals. Finally, the paper is concluded in Section IV.
\color{black}

\underline{\emph{Notations:}} In this paper, x, \textbf{x}, \textbf{X} denote a scalar, vector and matrix, respectively. Symbols $\textbf{x}^{T}$  and  $\textbf{X}^{T}$ represent transpose of \textbf{x} and \textbf{X}, respectively. Symbols $\mathbb{B}$ and $\mathbb{C}$ %, $\mathbb{Z}$ and $\mathbb{R}$
represents the set of binary numbers and complex numbers, %integers and real numbers, 
respectively. Also, the ${i}$th element of vector $\textbf{x}$ is denoted by $\textbf{x}_i$ and $\text{diag(\textbf{x})}$ denotes a diagonal matrix in which $i$th diagonal element is $\textbf{x}_i$.   The maximum value of $f(x)$ as $x$ is varied over all its possible values is denoted by `$\max \limits_{x} f(x)$'. 
\color{black}
$ x \in \mathcal{CN} (\mu,\sigma^2)$ denotes that $x$ is a circularly complex Gaussian random variable (RV) with mean $\mu$ and variance $\sigma^2$. \color{black}
The trace of a square matrix $\textbf{A}$ is denoted by $\text{Tr}(\textbf{A})$  which is calculated by the sum of main diagonal elements of $\textbf{A}$. %The selection of $r$ elements from a set of $n$ elements is denoted by %$\comb{n}{r}$ 
%$n\choose r$  
%$\binom{n}{r}$
%$\genfrac(){0pt}{2}{n}{r}$
The L2 norm of $\textbf{x}$ is denoted by  $\norm{\textbf{x}}$, and `log(\textbf{x})' denotes the natural logarithm of each element of $\textbf{x}$. $\mathcal{O}(\cdot)$ denotes
the complexity order. %$N \times M$ matrix denotes a matrix with $N$ rows and $M$ columns and $\textbf{I}_N$ denotes a  N × N identity matrix, respectively. 

\section{ System Model and SCMA Encoding}
In this section, we discuss the basic principles of SCMA and  the CB design, as it is one of the important issues of SCMA that affects the performance of the overall system.

\subsection{System Model}
Consider an  uplink  SCMA system in which $J$ users transmit data to the basestation (BS) using $K$ REs (e.g., time- or frequency- slots). In SCMA, the data/input bits of user is mapped to a complex codeword using the SCMA encoder. {For instance, user $j$ intends  to transmit $B$ coded bits $\textbf{b}_j=[b_j^1,\cdots, b_j^B] $}, \color{black} where $B=\text{log}_2(M)$ and $M$ is the modulation order. In SCMA, there is a CB defined for each user of size $K\times M$, and each column of this CB is called a codeword. Thus, for $J$ users, the combined set of the CBs for all the users is denoted by $ \Gamma$ of size $K \times M \times J$.  The SCMA encoder maps  the input bits of symbol $m, (m =1,\cdots, M)$ of the user $j$ to a codeword $\textbf{x}_j^m= \Gamma(:,m,j) = \mathbb{C}^{K \times 1}$.

\color{black}  Let \color{black} $\textbf{H}_j =\text{diag}[h_{1,j}, \cdots, h_{K,j}]^T$ be the effective channel matrix for the $j$th user, \color{black} where $h_{k,j}$ denotes the channel fading coefficient at the $k$th RE for the $j$th user. % Let %scmpot_156  $\textbf{x}_j=[x_{1,j},\cdots,x_{K,j}]^T$ be the transmitted codeword of the $j$th user, where $x_{k,j}$ be the codeword element transmitted by the $j$th user on the $k$th RE.
 The received signal at the BS is 
\begin{align}
    \textbf{y} = \sum_{j=1}^{J}  \textbf{H}_{j} \textbf{x}_{j}^m + \textbf{n},
    \end{align}
where   $\textbf{n}$ $\in$ $\mathbb{C}^{K \times 1}$ is the noise vector, each element of which is modeled by complex Gaussian distribution $\mathcal{CN} (0,\sigma^2)$. 
Due to the sparse nature of SCMA CBs,  $d_f$ out of $J$  users overlap over each RE, and  each user data is transmitted on $d_v<K$ REs.  Let $\xi_k$ be the set of users transmitting data over the $k$th RE and $\zeta_j$ be the set of REs on which the $j$th user has active transmission. By definition, the cardinalities of   $\xi_k$ and $\zeta_j$ are $d_f$ and $d_v$, respectively. Thus, the received signal at the $k$th RE is   given as
\begin{align}
        % y_k=\sum_{j_i \in N(k)}^{} h_{kj_i} C_{k,j_i}(\textbf{x}_{j_i}) +n_k, \hspace{1cm}     \text{for}~k=1,2,\cdots, K.
        %
        y_k=\sum_{j \in \xi_k}^{} h_{k,j} x_{k,j}^m +n_k, ~     \text{for}~k=1,2,\cdots, K,
\end{align}
 where  $x_{k,j}^m$ denotes the codeword element  transmitted by user $j$ over the  $k$th RE corresponding to symbol $m$, and $n_k$ is the noise received at the $k$th RE. \color{black}In this article, superscript $m$ is neglected sometimes, in case the modulation symbol does not need to be specified.
 \color{black}\\
$-$ \emph{Example:}
 Let us consider  $4 \times 6$ SCMA block encoding having $K=4$ REs and $J=6$ users as shown in Fig. \ref{fig:encode} with overloading factor $\lambda=J/K=1.5$. The allotment of REs  among users can  be represented by a factor graph matrix, where   each row corresponds to a RE, and each column corresponds to a user, respectively. The number of ones in a row is $d_f$, and the number of ones in a column is $d_v$. $\textbf{F}_{kj}=1$ denotes that user $j$ has active transmission over the  $k$th RE. For a $4 \times6$ SCMA block, the factor graph matrix of order  $4 \times 6$ corresponding to Fig. \ref{fig:encode} is given below:

 \begin{align}\label{fac_gra_4_6}
    \textbf{F}_{4 \times 6}=
    \begin{bmatrix}
    1&1&1&0&0&0\\
    1&0&0&1&1&0\\
    0&1&0&1&0&1\\
    0&0&1&0&1&1\\
    \end{bmatrix},
\end{align}

\begin{figure*}[htbp]
\centering
\includegraphics[scale=0.8,trim=1cm 5cm 1.5cm 11cm,clip]{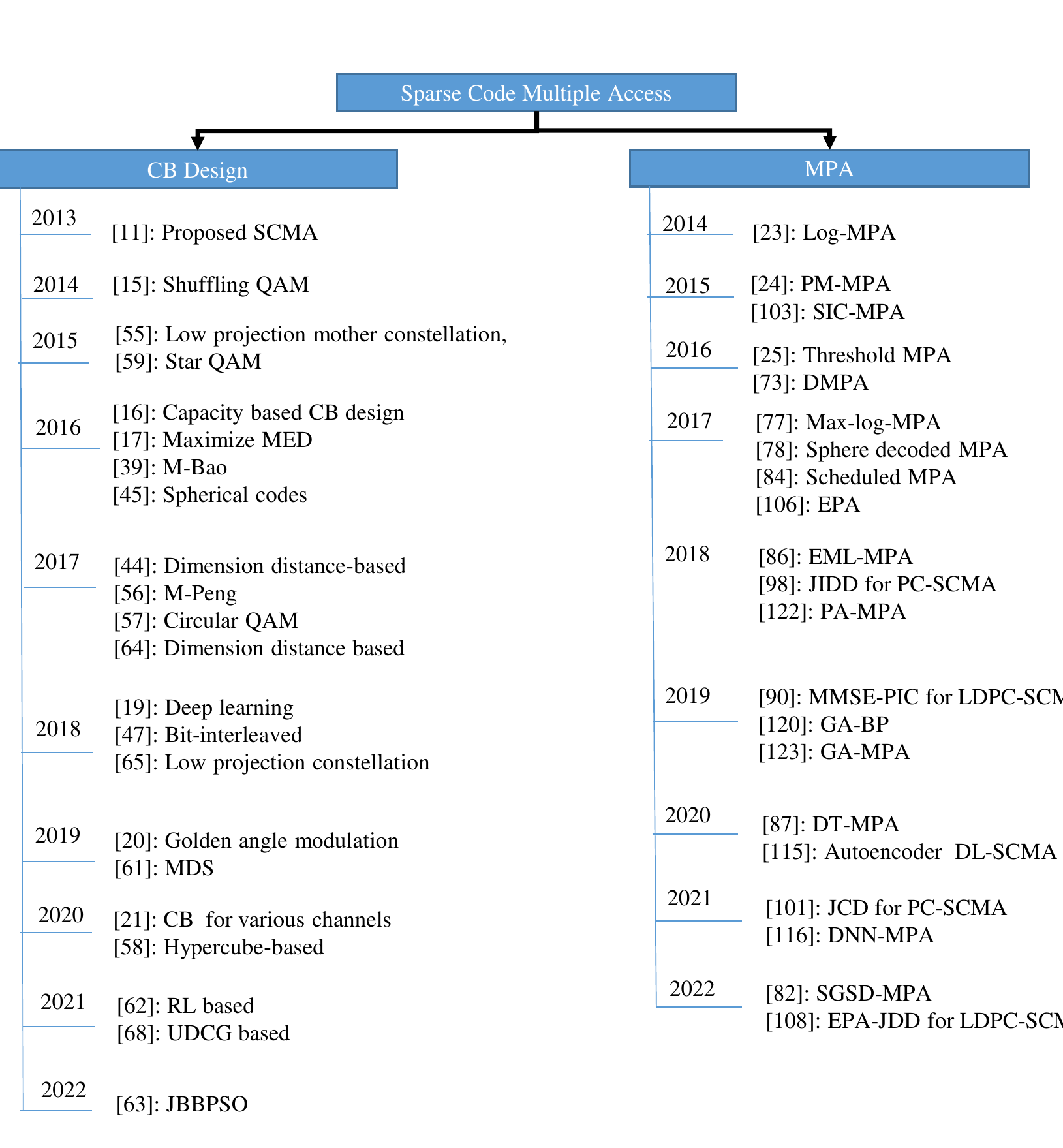}
\caption{Development of SCMA.}
\label{fig:scma_development}
\end{figure*}

 Note that the number of users superimposing over one  RE is 3, i.e., $d_f=3$ and the number of non-zero values in every column is 2, i.e., $d_v=2$. For example, the  received signal at the second  RE can be written as 
\begin{align}
            y_2  =h_{2,1}  x_{2,1}+h_{2,4}  x_{2,4} +h_{2,5}  x_{2,5} +n_2, %~~~~~~~~~~~~~~~~~~~~~~~~~\text{for}~k=1,2,\cdots, K.
\end{align}
where ${x}_{2,1},{x}_{2,4},{x}_{2,5}$ are the codeword elements of the first, fourth and fifth users on second RE ($\xi_2=\{1,4,5\}$), respectively. 
  \color{black}Further, the discussion on SCMA is mainly divided into two areas, namely CB design and MPA for decoding. Fig. \ref{fig:scma_development} shows
 a timeline of the development of SCMA CB design and decoding
algorithms in  chronological order.

 \color{black}

\subsection{SCMA Encoding and CB Design}

%Using a well designed CB (CB) can significantly increase the spectrum efficiency as well as reduce the detection complexity. 
The performance gain of SCMA over other NOMA schemes is strongly dependent on judiciously  designed sparse CBs.  The CB of each user has its own sparsity pattern.% and can be written as a matrix of size $K\times M$, where $M=2^B$. %denotes the number of codewords of each user. %In every specific  CB, each column vector (i.e., a codeword) is sparse consisting of  $d_v$ non-zero elements at certain fixed resource elements (REs) pertinent to a specific user.
The SCMA
CB design problem involves finding the optimum  mapping matrix, $\textbf{V}^*$, along with the optimum multidimensional
constellation, $\textbf{A}^*$, and may be defined as
\begin{align}\label{cb_opt}
    \textbf{V}^*, \textbf{A}^* = \mathop{\arg \max}_{\textbf{V},\textbf{A}} D(S(\textbf{F},\textbf{A}; J,M,d_v,K))
\end{align}
where $D$ is the design criterion for the SCMA system, $S$, with $J$ users and $K$ REs.  
Albeit numerous families of  CBs have been proposed in the literature,  optimal CB design remains  open. The current designs are mostly sub-optimal and  are based on a  multi-stage approach \cite{scmapot_14_scma_1,cb2,uplinkbcsurvey_36, cb3, uplinkbcsurvey_38,uplinkbcsurvey_39,uplink_cb_survey_41,uplink_cb_survey_46,uplink_cb_survey_48, uplink_cb_survey_42, uplink_cb_survey_43, uplink_cb_survey_44,uplink_cb_survey_45, uplink_cb_survey_47 }. For the $j$th user, multi-dimensional CB, 
\color{black} $\text{CB}_j= \Gamma(:,:,j)$ of size $K \times M$\color{black} ~can be expressed as
\begin{align}\label{cb}
    \text{CB}_j=\textbf{V}_j \Delta_j \textbf{A}_{\text{MC}},~~ \text{for}~j=1,2,\cdots,J,
\end{align}
where $\textbf{V}_j \in \mathbb{B}^ {K \times d_v}$ denotes the binary mapping matrix, $\textbf{A}_j= \Delta_j \textbf{A}_{\text{MC} }$,
$\textbf{A}_{\text{MC}}$ denotes the multi-dimensional mother constellation   and $\Delta_j$ refers to the constellation operator for the $j$th user.
%\cite{cai2016multi}.  
%with the aid of multi-dimensional mother constellations, mapping matrices,  and user specific operations such as interleaving, permutations, and phase rotations. 
% Specifically, for  user $j$, bits $\textbf{b}_j$ are mapped to a complex codeword $\mathbf{m}_j$ using an SCMA encoder. 
%\begin{align}
    %    \textbf{x}_j = \textbf{V}_j (\Delta_j \circ \rho) \textbf{b}_j, ~~ \text{for}~ j=1,\cdots,J.
%\end{align}
%where $\mathbf{m}_j \in \mathbb{A} \hspace{0.1cm} \subset \hspace{0.1cm} \mathbb{C}^K$, $\mathbb{A}$ denotes the set of codewords allocated to user $j$, cardinality $|\mathbb{A}|=M$. $\textbf{V}_j$ \in \hspace{0.1cm} \mathbb{B}$^ {K \times N}$ denotes the binary mapping matrix of size $K\times N$, $\rho$ denotes the mother constellation   and $\Delta_j$ refers to the constellation operation for the $j$th user \cite{cai2016multi}.  
The mapping matrix is selected so that each user has active transmissions over a few fixed REs only. 
Each column of the factor graph % matrix $\textbf{F}=[\textbf{f}_1,\textbf{f}_2,\cdots,\textbf{f}_J]$, where 
$\textbf{f}_j=\text{diag}(\textbf{V}_j \textbf{V}_j^T)$. The binary mapping matrices of six users corresponding to the factor graph matrix in (\ref{fac_gra_4_6}) are given below
\begin{equation*}
\begin{split}
    \textbf{V}_1 =
\begin{bmatrix}
1 & 0\\
0&1\\
0 & 0\\
0 &0\\
\end{bmatrix}, 
\textbf{V}_2 =
\begin{bmatrix}
1 & 0\\
0 & 0\\
0 & 1\\
0 & 0\\
\end{bmatrix}, 
\textbf{V}_3 =
\begin{bmatrix}
1 & 0\\
0 & 0\\
0 & 0\\
0 & 1\\
\end{bmatrix}, 
\\\\
\textbf{V}_4 =
\begin{bmatrix}
0 & 0\\
1 & 0\\
0&1\\
0 &0\\
\end{bmatrix},
 \textbf{V}_5 =
\begin{bmatrix}
0 & 0\\
1 & 0\\
0 & 0\\
0&1\\
\end{bmatrix},
\textbf{V}_6 =
\begin{bmatrix}
 0&0\\
0 & 0\\
1 &0\\
0&1\\
\end{bmatrix}.
\end{split}
\end{equation*}
$\textbf{V}_1$ indicates that user 1's data is transmitted over the first and the second REs. Similarly, $\textbf{V}_2$ indicates that user 2's data is transmitted over the first and the  third REs, and so on. Once the mapping matrices are decided, the next step is to design the mother constellation $\textbf{A}_\text{MC}$ and constellation operator, $\Delta_j$. Next, we discuss  the key performance indicators
(KPIs) that should be considered while designing $M$-point multi-dimensional constellations.

\subsubsection{KPIs}
The KPIs of $M$-point multi-dimensional constellations that affect  the performance of SCMA systems  are as follows:
\begin{itemize}
    \item Euclidean Distance: The minimum Euclidean distance between two multi-dimensional
constellation points, $\textbf{x}^m$ and $\textbf{x}^{m'}$, $ 1\leq m <m'\leq  M$, is given as
\begin{align*}
    d_{E,min}= \text{min} \left\{\norm{\textbf{x}^m - \textbf{x}^{m'}} \vert  1\leq m <m'\leq  M  
    \right\}.
\end{align*}
It is known that $d_{E,min}$ is a KPI for designing the constellations in the case of AWGN channels \cite{uplink_cb_survey_58, uplink_cb_survey_56,uplink_cb_survey_57}. If each user observes the
same channel coefficients over their $d_v$ REs (dependent fading), the
relative Euclidean distance between the constellation points
remains the same. So, in this case, $d_{E,min}$ is an important factor in the design process of constellation points.

\item Euclidean Kissing Number ($\tau_E$): It is defined as the average number of constellation pairs at  distance $d_{E,min}$, and its value should be low for designing the constellation points in the case of dependent fading.

\item Product Distance: The minimum product distance between the constellation points is given as
\begin{align*}
    d_{P,min}= \text{min} \Big\{{\prod_{k \in \mathcal{K}_{mm'}}{ \vert {x}_{k}^m - {x}_{k}^{m'} } \vert  
    }\Big\},  \\     ~~~~  \text{for}~ 1\leq m <m'\leq  M,
\end{align*}
where  $x_{k}^m$ and $x_{k }^{m'}$ are the $k$-th codeword elements of $\textbf{x}^m$ and $\textbf{x}^{m'}$, and 
$\mathcal{K}_{mm'}$ denotes the set of
dimensions, $k$, for which $x_{k}^m \neq  x_{k}^{m'}$. It is shown that the minimum
product distance of the constellation points should be maximized to have a positive
impact on the performance of the SCMA system, especially when each user observes different
channel coefficients over their $d_v$ REs (independent fading) \cite{uplink_cb_survey_58,uplink_cb_survey_59,uplink_cb_survey_60}.

\item  Product kissing number ($\tau_P$): 
It is defined as the  number of constellation pairs
at the minimum product distance. The product kissing number should be low to attain low error probability, especially in the case of independent fading \cite{uplink_cb_survey_59}. 

\item Diversity ($L$): The  diversity of the signal constellation 
 is the minimum number of different components
of any two distinct constellation points. $L$ should be high to attain a low error probability in case of independent fading. 

\end{itemize}

\subsubsection{Designing of Mother Constellation}
The mother constellation matrix $\textbf{A}_{\text{MC}} \in \mathbb{C}^ {d_v \times M}$  represent
each $\text{log}_2(M)$ bits with a codeword of $d_v$ elements. In general,  the mother constellation is designed to possess good distancing properties according to the KPIs mentioned above. Also, to mitigate multi-user interference,  dependency
among the nonzero codeword elements is induced such that the receiver can recover the  overlapping codewords from other REs. Several works were proposed  to optimize the design of the mother constellation \cite{uplinkbcsurvey_36,scmapot_78,scmapot_80, uplink_cb_survey_41,uplink_cb_survey_42,scmapot_90,scmapot_92,starqam,scmapot_76, uplink_cb_survey_44,scmapot_97}.

In \cite{cb2}, 
a mother constellation was designed such that the real and imaginary parts are independent, while the complex dimensions are still dependent. A shuffling approach
was used, which reduces the detection complexity from $M^{d_f}$
to $M^{d_f/2}$. \color{black}  In \cite{uplink_cb_survey_41}, constellation based on the rotation of QAM
constellations was proposed. The approach was known as M-Bao, and  feasible if $d_f \leq 3$. In \cite{scmapot_78},  a low-projection mother constellation was proposed
that  employs  lesser number of colliding
constellation points over each RE, and also reduces the detection complexity. In \cite{scmapot_76}, the M-Peng scheme is  
designed by maximizing the minimum Euclidean distance between the constellation points of all the users.   
\color{black} In \cite{starqam,scmapot_80, scmapot_92}, constellation points of a given dimension in the mother constellation  belong to the  concentric rings. The codeword elements for the  first dimension of the mother constellation are taken from  a  star-QAM constellation, then the codeword elements of
other dimensions are generated by scaling and permuting the
points of the first dimension. The  objective
 was to minimize the pairwise error probability (PEP)
between two superimposed  codewords $\textbf{s}_a, \textbf{s}_b$ which is given by
\begin{align}
    \mathbb{P}(\textbf{s}_a, \textbf{s}_b \vert \textbf{H})= Q (\sqrt{\frac{\norm{\textbf{H} (\textbf{s}_a -\textbf{s}_b)  
    }^2}{2N_0}})
\end{align}
where $N_0$ is the noise variance. This approach is powerful because the PEP criteria  is applied directly to the  codewords of  $\textbf{CB}_j$, and this method can be extended  for generating large CBs. 
\color{black}
A large CB refers to CB with a large $M,J,K,d_v$. For example, CB with $M=16,K=6,J=12,d_v=3$ is said to be large.
%Large CBs mean that modulation order can be  $M=4,8,16$, number of REs, $K=4,5,6,8$ and number of non-zero values in a codeword, $d_v=2,3,4$. 

\color{black}
In some works, the approach was to solve the optimization problem defined in (\ref{cb_opt}) to generate the CBs. The problem in (\ref{cb_opt}) is usually
%non-convex quadratically constrained quadratic programming problem \cite{uplink_cb_survey_47}, which is
NP-hard and  is  difficult to solve\cite{uplink_cb_survey_47}. Thus, the optimization problem is reformulated and solved using different optimization techniques\cite{uplink_cb_survey_42,scmapot_76,scmapot_90,scmapot_97}. \color{black} In \cite{2019_1}, CB design 
using maximum distance separable (MDS) codes was proposed
which provides improved signal-space diversity. In \cite{2021_2}, CB design based on Reinforcement Learning (RL) was proposed which are 
suitable for  large-scale SCMA
schemes. In \cite{cb_2022_1},  a novel CB optimization method, namely  joint bare bones particle swarm optimization
(JBBPSO) was proposed to  maximize the average mutual information.
\color{black}

%Unitary rotations may be applied to a mother constellation to increase power variation among different users in order to reinforce the ``near-far effect" for suppression/mitigation of multiuser interference and to enhance the constellation shaping gain. 
\subsubsection{Constellation Operators}
% constelltion opertor name I have taken from sta qaam paaper.
\color{black}
Once the mother constellation is designed, constellation operators, $\Delta_j$ are applied to generate multiple CBs for different users. The operators should be
designed to mitigate the multi-user interference and increase the decodability of the multi-user signals at the
receiver.
A constellation operator, $\Delta_j$ (as shown in (\ref{cb})) may include either one or combination of phase rotation, complex conjugate, user power offset, and dimensional permutation\cite{starqam, scmapot_98,uplinkbcsurvey_39,scmapot_100,scmapot_99}. The main motive of applying constellation operators 
is to maintain uniquely decodable symbols for users to 
collide at the same RE. %The effect of the constellation operators is mainly in the downlink scenario. In the case of uplink, the users experience different fading channels.
%uplink cb suvrey paper
In \cite{scmapot_99}, the optimization of the constellation operator was proposed
 to find the most appropriate
permutation set, which has the maximum sum of the distances between the interfering codewords. In \cite{scmapot_94}, a multistage optimization scheme was proposed to design
low-projected multidimensional constellations.
A uniquely decomposable constellation group
(UDCG) based
CB  design was proposed in \cite{udcg_cb} for larger modulation order.
\color{black}
It should be noticed that the designing criteria $D$ in  (\ref{cb_opt}) should not be compromised with the constellation operators.
%\fbox{\includegraphics[scale=0.4,trim=12cm 4.5cm 10cm 3.5cm,clip]{star_qam_rings.pdf}}
%trim=left bottom right top
\begin{figure}
\centering
\includegraphics[scale=0.4,trim=12cm 4.5cm 10cm 3.5cm,clip]{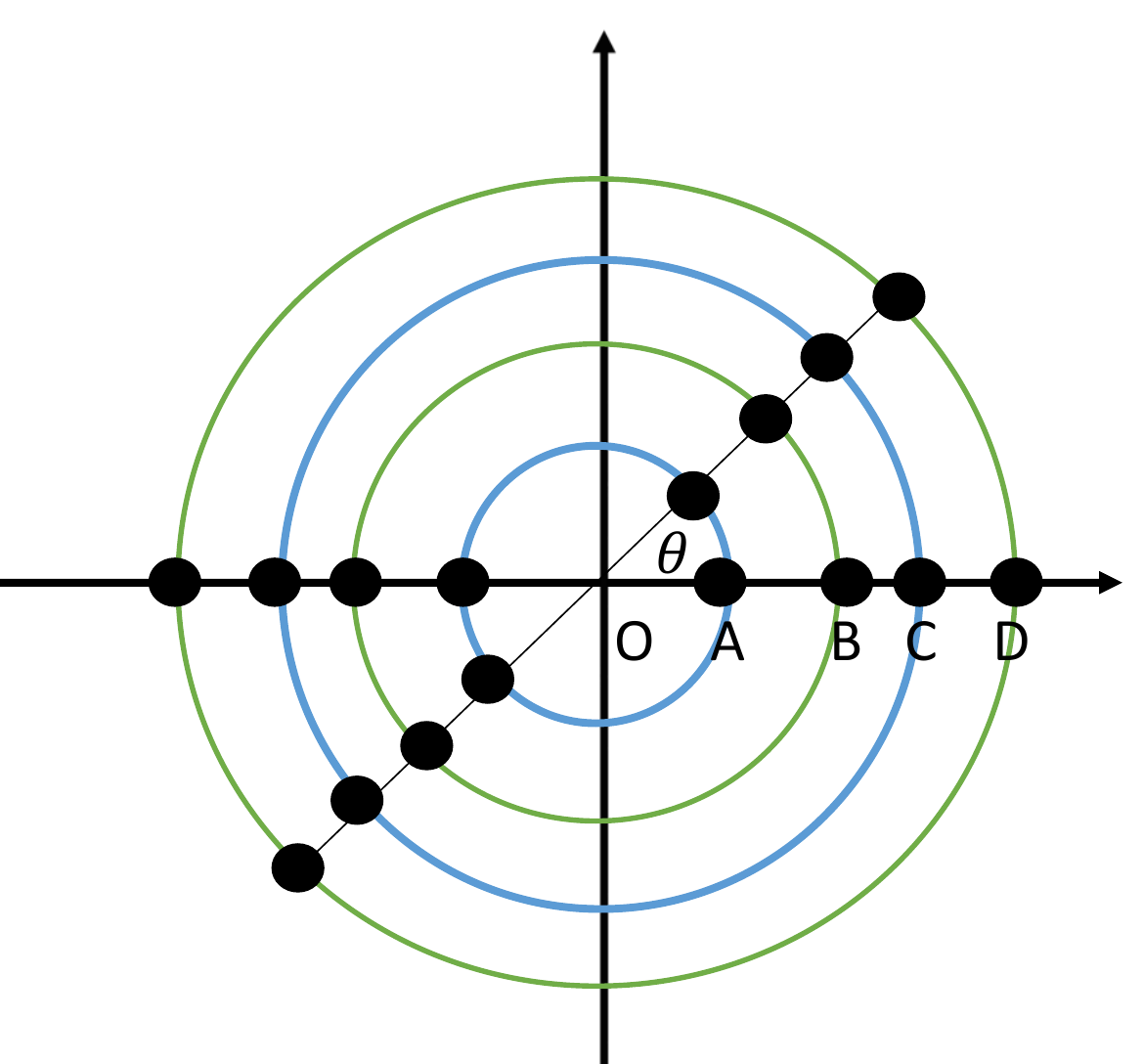}\\
\caption{Four-rings star-QAM mother constellation with $d_v=2 $ and $M = 4$.}
\label{fig:starqam_rings}
\end{figure}
 
%It is generally observed that higher the power diversity among interfering users, it becomes easy to decode interfering codewords at the receiver. 
Next,  an example of SCMA CB  design for $J=6, K=4,M=4,d_v=2$  based on $M$-dimensional star-QAM constellations is provided \cite{starqam}:
\begin{itemize}
    
\item{Step 1: Designing the mother constellation, $\textbf{A}_{MC}$: } 
The mother constellation is designed based on the four concentric rings star-QAM signaling constellation, as shown in Fig. \ref{fig:starqam_rings}. 
\begin{align*}
|OA| = R_1, |OB| = R_2, |OC| = R_1^{'}, |OD| = R_2^{'},    
\end{align*}
where $\alpha=\frac{R_2^{'}}{R_2}=\frac{R_1^{'}}{R_1}$ and $\beta=\frac{R_2^{'}}{R_1^{'}}=\frac{R_2}{R_1}$.  Thus, the mother constellation,
$\textbf{A}_{MC}$, can be expressed as
\begin{align}
 \textbf{A}_{MC}=
 \begin{bmatrix}
    \alpha R_1 & R_{1} & -R_{1} & -\alpha R_{1}\\
    -R_{2}& \alpha R_{2} & -\alpha R_2 & R_{2}\\
        \end{bmatrix}
    {.}
\end{align}
where $R_1= \sqrt{\frac{1}{2(\alpha^2+\beta^2+\alpha^2 \beta^2+1 )}}$, assuming  each user’s
CB energy is equal to one, and the value of $\alpha=3, \beta=1/0.62$ \cite{starqam}.

\item{Step 2: Constellation operators:}
The constellation operators for six users are as follows:
\begin{equation*}
\begin{split}
    \Delta_1 =
\begin{bmatrix}
e^{i \theta_1} & 0\\
0& e^{i \theta_2}\\
\end{bmatrix}, 
\Delta_2 =
\begin{bmatrix}
1 & 0\\
0 & 1\\
\end{bmatrix}, 
\\\\
\Delta_3 =
\begin{bmatrix}
0 & e^{i \theta_3}\\
e^{i \theta_1} & 0\\
\end{bmatrix}, 
\Delta_4 =
\begin{bmatrix}
1 & 0\\
0 & e^{i \theta_2}\\
\end{bmatrix},
\\\\
 \Delta_5 =
\begin{bmatrix}
0 & 1\\
1 & 0\\
\end{bmatrix},
\Delta_6 =
\begin{bmatrix}
 1&0\\
0 & e^{i \theta_3}\\
\end{bmatrix}.
\end{split}
\end{equation*}
Traditionally, $\theta_1=0,\theta_2= \pi/3$ and $ \theta_3= \frac{2 \pi}{3}$.
Then, the CB for first user is $ \textbf{CB}_1=\textbf{V}_1 \Delta_1 \textbf{A}_{\text{MC}}$, and similarly CB for other users can be generated using (\ref{cb}).
 \end{itemize}
Above CB construction method based on star-QAM constellation can be extended for large size or higher dimension CB generation \cite{scmapot_92}. 

\section{Decoding}
The process of extracting  data from the noise corrupted received 
signal  is called decoding, and there are two types of decoding, namely hard and soft decoding.
%\subsection{Soft Decoding}
In some communication systems, conventional hard decoding is  used, where each decision is taken at the receiver based on a certain threshold value. For instance, a simple hard decoding receiver decides that if  the received value is greater than the threshold value, it will be decoded as 1, and 0 otherwise. However, it is not verified how close or far is the received value to the threshold before making the decision. An alternative approach is to obtain the estimated sequence and its reliability level, where the latter indicates the `confidence' we have in that estimated sequence. Such a decoding approach is called soft decoding \cite{book_soft_iter, siso_book2}. In this way, more information can be extracted from the received data, and estimation/detection can be carried out in an iterative and  improved manner. 

\subsection{MAP Detection}
The objective of an optimal detector is to minimize the probability of error $(P(e))$ for the transmitted bit sequence, i.e., to minimize the mismatch between transmitted bits ($\textbf{b}$) and estimated bits ($\hat{\textbf{b}}$).
\begin{align}\label{prob_err}
    \min P(e)= \min P(\textbf{b} \neq \hat{\textbf{b}})\,.
\end{align}
Given all the possible observations  \textbf{y} and the channel coefficient matrix \textbf{H}, we have
\begin{align}\label{error}
    P(e)= \int P(e \vert \textbf{y}) f(\textbf{y}) d\textbf{y}\,,
\end{align}
where $P(e \vert \textbf{y})$ indicates the probability of error given received vector $\textbf{y}$ and $f(\textbf{y})$ indicates the probability density function (PDF) of \textbf{y}. It is noted from (\ref{error})  that $P(⁡e)$ is proportional to $P(e \vert \textbf{y})$. 
Since  $\textbf{b}$ is the desired data,
\begin{align}
    P(e \vert \textbf{y})=1-P(\textbf{b} \vert \textbf{y})\,,
\end{align}
where $P(\textbf{b} \vert \textbf{y})$ denotes the probability of $\textbf{b}$ given the vector $\textbf{y}$. Thus, to minimize the probability of error, $P(\textbf{b} \vert \textbf{y}) $ needs to be maximized and this detection scheme is termed as MAP detection. Using Bayes’ rule,
\begin{align}
    P( \textbf{b} \vert \textbf{y})= \frac{f(\textbf{y} \vert \textbf{b})P(\textbf{b})}  {f (\textbf{y}) }\,,
\end{align}
 where $f(\textbf{y} \vert \textbf{b})$ indicates the conditional PDF  of $\textbf{y}$ given $\textbf{b}$ and $f(\textbf{y})$ indicates the PDF of $\textbf{y}$. %Since $f(\textbf{y})$ is constant for all the values that $\textbf{s}$ can take and 
 Assuming the prior probabilities  $P(\mathbf{b})$ are identical for different $\mathbf{b}$, so
 \begin{align}
     P(\textbf{b} \vert \textbf{y}) \propto f(\textbf{y} \vert \textbf{b} )\,.
 \end{align}
 Thus, the $ \textbf{b} $ which  maximizes $f(\textbf{y} \vert \textbf{b} )$ is chosen to minimize the probability of error and this detection scheme  is called the maximum likelihood (ML) detection. In short, the MAP detection is equivalent to the ML detection when prior probabilities take the identical value.\\\\
\textbf{MAP Detection in SCMA:}\\
For an SCMA decoder, given \textbf{y} and the channel matrix \textbf{H} at the receiver, the detected multiuser  codeword $\hat{\textbf{X}}=[\textbf{x}_1,\textbf{x}_2,\cdots,\textbf{x}_j]$ is given as
\begin{align} \label{multiuser_codeword}
    \hat{\textbf{X}} = \argmax_{\textbf{x}_j \in \textbf{CB}_j,\forall j} p(\textbf{X} \vert \textbf{y})\,,
\end{align}
%where $\textbf{x}_j$ represents the codeword transmitted by the $j${th} user and $ \mathbb{A}_j$ represents the set of codewords allotted to the $j$th user. 
The transmitted codeword of each user can be estimated by maximizing
its a-posteriori probability mass function (PMF). This can be calculated by taking
the marginal of the joint a-posteriori PMF defined in (\ref{multiuser_codeword}). %So, the objective is to detect the codeword transmitted by each user one by one.
\begin{enumerate}
    \item [$\circ$]  \textbf{{Marginalization:}} It is known that  if $\mathcal{V}$ is a function of several variables, then $\mathcal{V}$ with respect to a specific variable can be obtained by carrying out marginalization with respect to that variable.

\begin{itemize}
    \item [$-$] \emph{Example:} Let $\mathcal{V}(\alpha_1,\alpha_2, \cdots, \alpha_j,\cdots, \alpha_J)$ be a function (also called a global function) of $J$ number of variables, then the marginalized function with respect to $\alpha_j$ variable is given as
\begin{align}
    \mathcal{V}(\alpha_j)= \sum_{\alpha_1} \cdots \sum_{\alpha_{j-1}} \sum_{\alpha_{j+1}} \cdots \sum_{\alpha_J} \mathcal{V}(\alpha_1, \cdots, \alpha_J)\,,
\end{align}
i.e., the summation with respect to all the variables except  $\alpha_j$. It can be written   in a more compact form as
\begin{align}\label{marginalization}
    \mathcal{V}(\alpha_j)= \sum_{{\sim}\alpha_j}  \mathcal{V}(\alpha_1,\alpha_2, \cdots, \alpha_j,\cdots, \alpha_J)\,,
\end{align}
where $\sim\alpha_j$ denotes that the summation is performed with respect to all the variables of the global function except for $\alpha_j$.

\end{itemize}
\end{enumerate}
Similar to (\ref{marginalization}), the marginalization with respect to $\textbf{x}_j$ in  $(\ref{multiuser_codeword})$ leads us to symbol by symbol detection for the  $j$th user which can be given as
\begin{align}
    \hat{\textbf{x}}_j = \argmax_{\textbf{x}_j \in \textbf{CB}_j} \sum_{{\sim} \textbf{x}_j} p(\textbf{X} \vert \textbf{y})\,.
    %  p is pmf function.
\end{align}
Again using Bayes’ rule, we have
\begin{align}\label{map}
    \hat{\textbf{x}}_j = \argmax_{\textbf{x}_j \in \textbf{CB}_j} \sum_{{\sim} \textbf{x}_j} f(\textbf{y} \vert \textbf{X})  P(\textbf{X})\,.
\end{align}
For SCMA system, assume that the  noise components corresponding to the $K$ REs are identically distributed and independent of each other. As a result, we have 
\begin{align}
    f(\textbf{y} \vert \textbf{X})= \prod_{k=1}^{K} f(y_k \vert \textbf{X})\,,
\end{align}
where $k=1,\cdots,K$ denotes the REs and $y_k$ is the received signal at the $k$th RE. Consequently,  (\ref{map}) becomes
\begin{align}\label{mpf}\nonumber
    \hat{\textbf{x}}_j = \argmax_{\textbf{x}_j \in \textbf{CB}_j} \sum_{{\sim}\textbf{x}_j} \left(P(\textbf{X}) \prod_{k=1}^{K} f(y_k \vert \textbf{X})\right)\,, 
    \enspace \\\text{for}~ j=1, \cdots, J.
\end{align}
Solving the marginal  product of functions (MPF) problem in  (\ref{mpf}) with brute force will lead to prohibitively high  complexity and hence may be infeasible when the number of users is large. Thanks to the sparsity of SCMA CBs, we will show in the sequel that low-complexity MPA can be leveraged with the aid of $\emph{factor ~graph}$ to solve the MPF problem in  (\ref{mpf}) with near-optimal performance.  

\subsection{Factor Graphs}
 Different types of graphs are used to model problems in areas such as computer science, biology, physics,  etc. A bipartite graph is widely known for modeling communication and signal processing problems. In this graph, total nodes can be divided into two sets, and no two nodes within a set are connected.

A \emph{factor graph} is an undirected bipartite graph in which one set of nodes is called variable nodes (VNs), and the other is called function nodes (FNs). An edge is connected between a VN and a FN if that particular variable is an argument of that function. The factor graph shows how a global function can be  represented in  simpler local functions (denoted by FNs) and can also help in computing marginal distribution with respect to a single variable using MPA.
%\begin{itemize}
    
%\item [$-$]  \emph{Example:} Joint distribution of two variables $(v_a, v_b)$ can also be simplified in conditional and prior probability distribution functions.
%\begin{equation} \label{eq1}
%\begin{split}
%f(v_a,v_b) & =f_1(v_a \vert v_b)f_2(v_b)\ \\
 %& = f_1' (v_a) f_2 (v_b)\,.
%\end{split}
%\end{equation}

\begin{figure}[h!]
\centering
\includegraphics[scale=0.3,trim=3cm 4.5cm 5cm 1.75cm,clip]{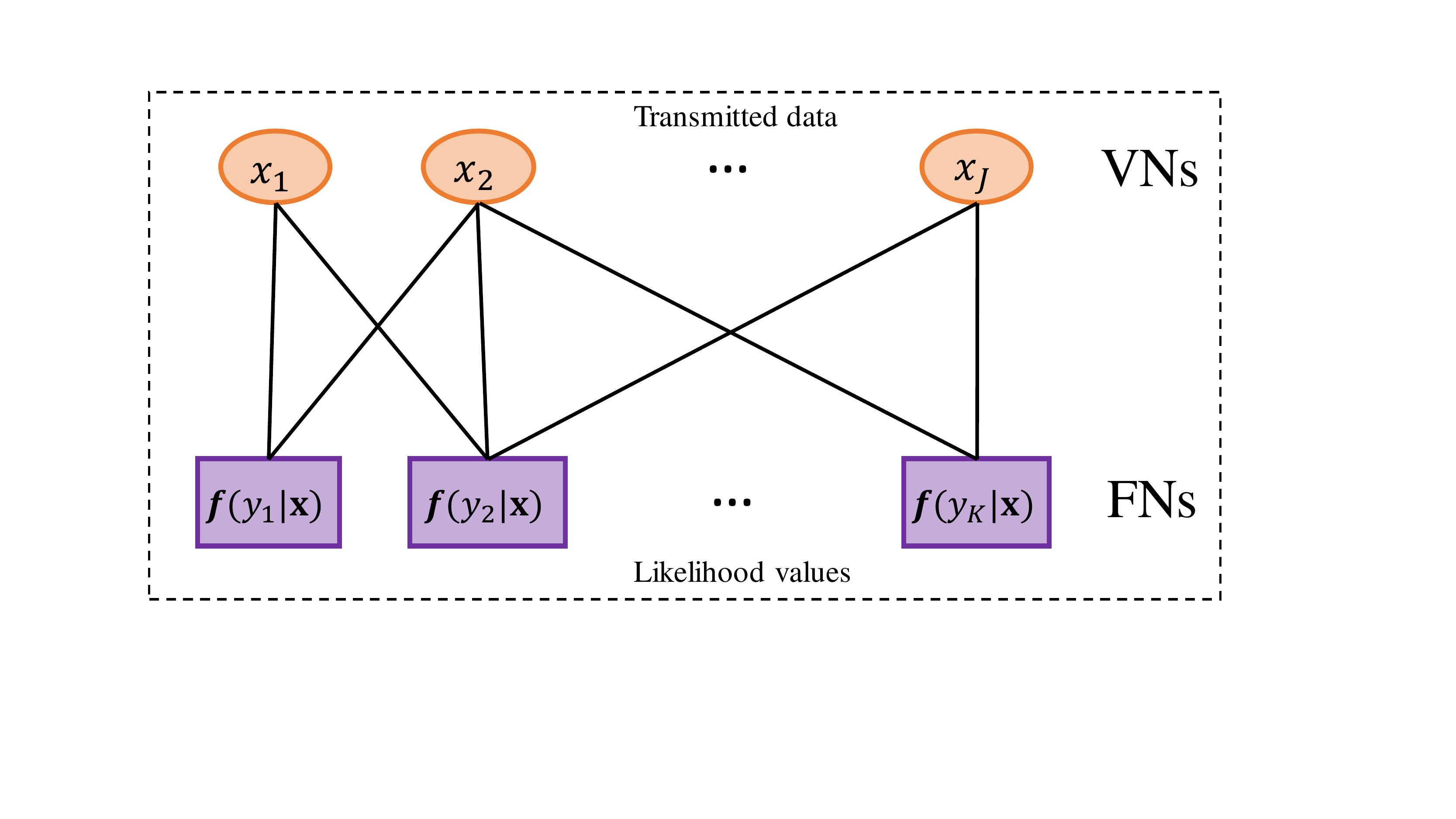}
\caption{Factor graph of a communication channel.}
\label{fig:fac_comm_ex}
\end{figure}

%\end{itemize}
Factor graphs are also widely used to model several communication problems. Fig. \ref{fig:fac_comm_ex} shows the factor graph of a communication channel \cite{tarokh}.
%\fbox{\includegraphics[scale=0.4,trim=3cm 4.5cm 5cm 1.75cm,clip]{fact_comm_ex.pdf}}
%trim=left bottom right top

Let $\textbf{x}=[x_1,x_2,\cdots, x_j,\cdots,x_J]$ be the transmit  vector and $\textbf{y}=[y_1 , y_2, \cdots, y_k, \cdots,y_K]$ be the received vector. Assuming  noise components are independent of each other, then
\begin{equation}
    f(\textbf{y} \vert \textbf{x})=\prod_{k=1}^{K} f(y_k \vert \textbf{x})\,.
\end{equation}
It is noted that each received symbol ($y_k$) depends on certain elements of \textbf{x}. Let say $y_k$ depends on $x_1,x_2,x_j$ elements, then there will be edge connecting FN corresponding to $f(y_k \vert x_1,x_2,x_j)$ with VNs $x_1,x_2,x_j$. In Fig. \ref{fig:fac_comm_ex}, every VN denotes the data from one transmit point and every FN $k$ denotes the likelihood function $f(y_k \vert \textbf{x})$. 

\begin{figure}[h!]
%\label{scma_fg}
\centering
\includegraphics[scale=0.32,trim=2.75cm 6cm 5cm 6cm,clip]{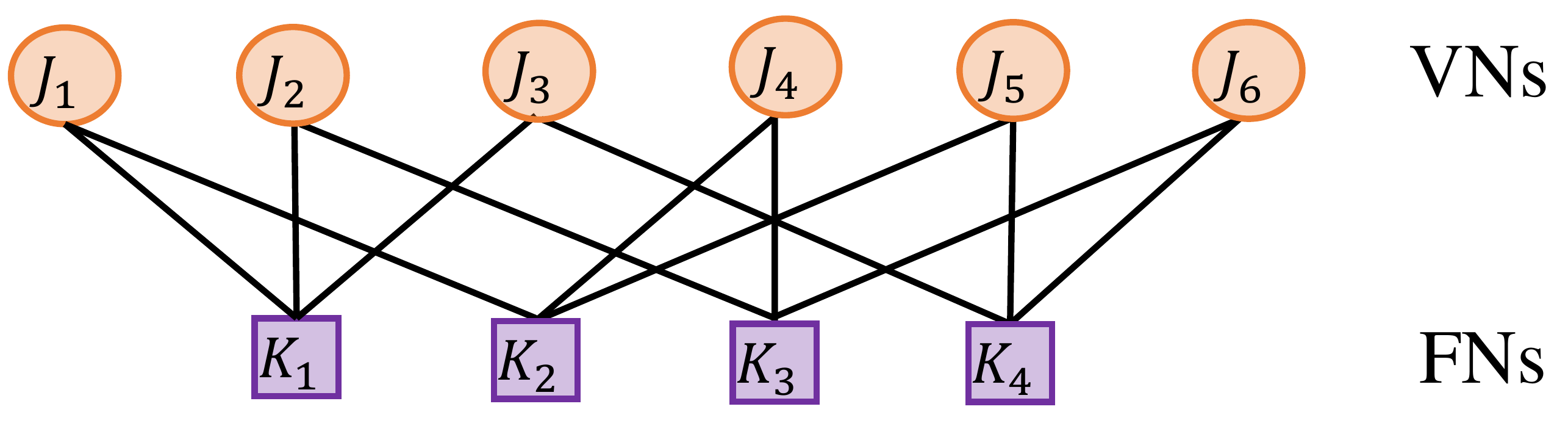}
\caption{Factor graph of the $4\times 6$ SCMA system corresponding to the factor graph matrix shown in (\ref{fac_gra_4_6}).}
\label{fig:scma_fac}
\end{figure}

%\fbox{\includegraphics[scale=0.5,trim=9.25cm 5.5cm 11.25cm 5.5cm,clip]{mpa_example_a.pdf}}
%trim=left bottom right top
%\fbox{\includegraphics[scale=0.5,trim=9.25cm 5.5cm 11.25cm 5.25cm,clip]{mpa_example_b.pdf}}
\begin{figure*}[h!]
    \centering
    \subfigure[]{    \includegraphics[scale=0.45,trim=9.25cm 5.5cm 11.25cm 5.5cm,clip]{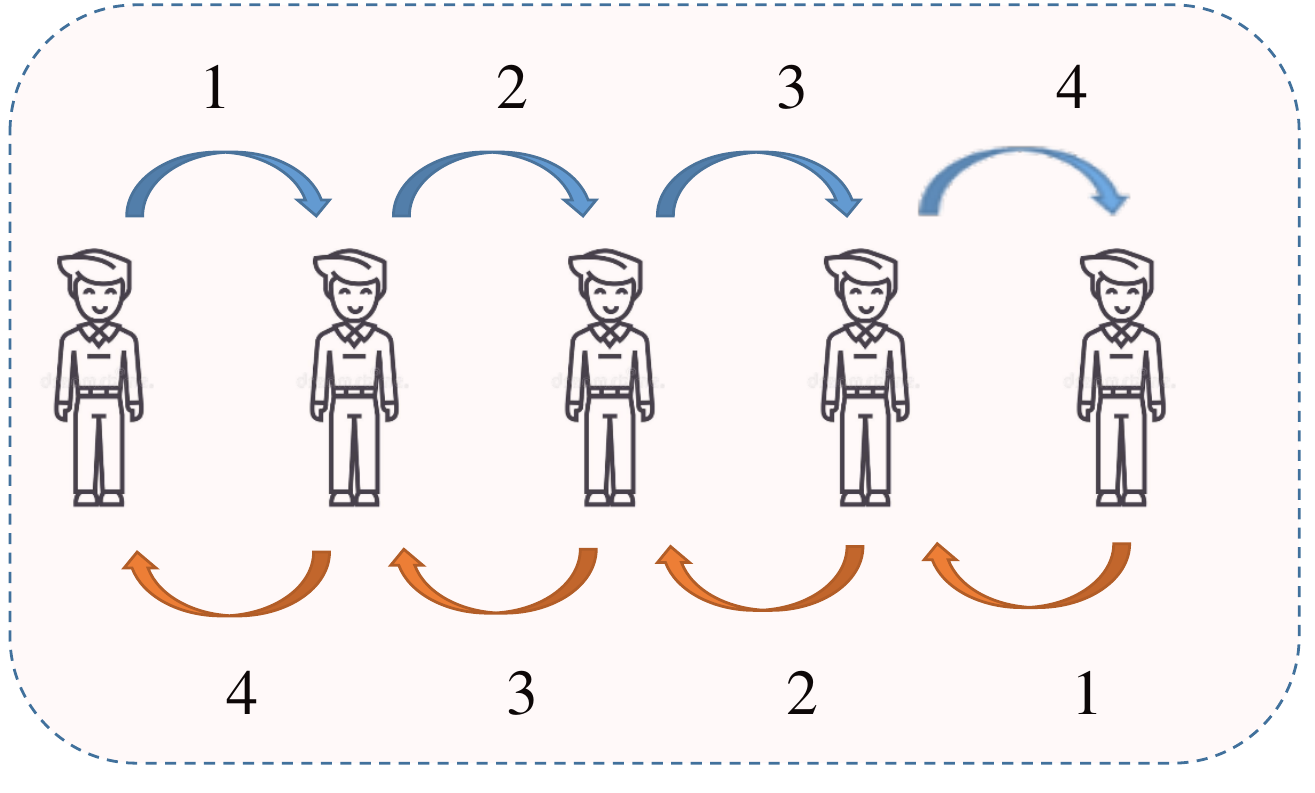}}
    %\label{fig:Mpa_ex_a}
    \subfigure[]{
    \includegraphics[scale=0.45,trim=9.25cm 5.5cm 11.25cm 5.25cm,clip]{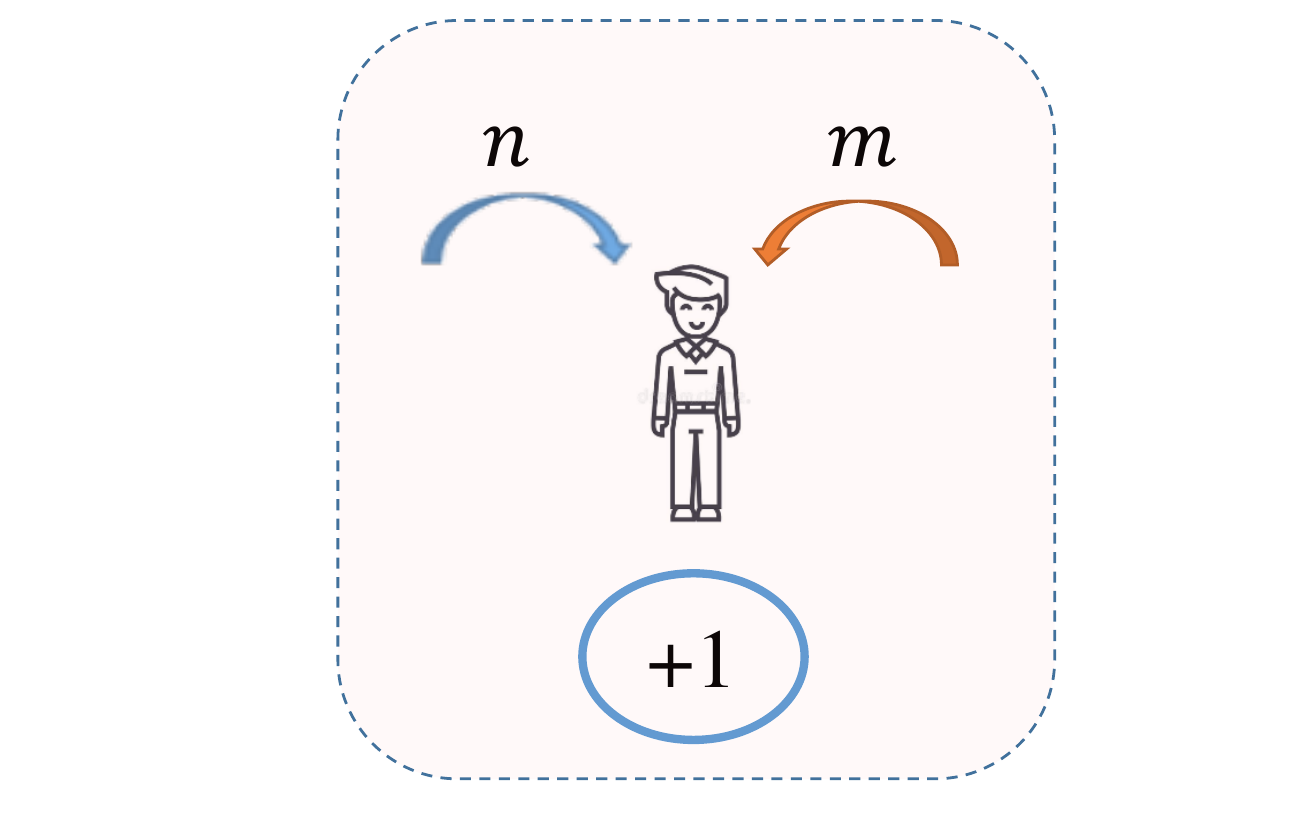}}
    %\label{fig:Mpa_ex_b}
    \caption{Students counting using MPA.}
    \label{fig:Mpa_ex}
    \end{figure*}
The factor graph corresponding to $4 \times 6$ SCMA system given in (\ref{fac_gra_4_6}) is shown in Fig. \ref{fig:scma_fac}. The first column of $\textbf{F}_{4\times6}$ indicates the first user, and it has non-zero values in the first and second rows. This means that data of the first user is transmitted on the first and second RE, so there is an edge between VN $J_1$ and FN $K_1$, and VN $J_1$ and FN $K_2$ (as shown in Fig. \ref{fig:scma_fac}). The second row of $\textbf{F}_{4\times6}$ indicates the second RE, and it has non-zero values at the first, fourth and fifth positions. This means that data of the first, fourth and fifth user overlaps on the second RE, and thus, there are edges connecting the FN $K_2$ with VN $J_1$, with VN $J_4$, and with VN $J_5$, respectively. %\ref{scma_fg}.

\subsection{Message Passing Algorithm}
MPA is an algorithm to conduct inference from graphical models by passing belief messages between the nodes. The following example may  help  understand how inference can be obtained from a graph \cite{book_soft_iter}.
\begin{itemize}
    \item[$-$] \emph{Example:} Fig. \ref{fig:Mpa_ex} (a) shows  a group of students standing in a line. The objective is to count the total number of students and pass this information to each student. The constraint is that one student can communicate with a maximum of two  neighboring students (front and back) at a time. To solve this problem, the approach proposed by MPA is that when one student receives a count from one side, he/she adds 1 to it (to indicate the student's presence) and passes it to the other side. The algorithm starts from each end of the line where the counter starts from 1 and increases by 1 as the message passes through each student. Messages pass in both directions of the line. If a student receives a count of $n$ from one side and $m$ from another side, then the total students present are $n+m+1$  (as shown in Fig. \ref{fig:Mpa_ex} (b)). 
    In this way, every student obtains the information on the total number of students present in the line.
    Such  problems can also be modelled with the help of a graph, where each node represents a student, and every connecting edge represents a communication link between them.
\end{itemize}

%\subsection{Sum-Product Algorithm}
 Next, we discuss   how messages are passed across a factor graph, and the resultant inference obtained using  MPA. In SCMA systems, each VN denotes one data layer, and each FN denotes the likelihood function at the RE. Therefore, the total number of VNs  equals the total number of layers/users, and the total FNs equals the total REs present.
 Consider a factor graph having $K$ FNs and $J$ VNs. Message passing in a factor graph using MPA is an iterative process if the factor graph has cycles (closed loops) %\footnote{A cycle is a path that starts from a node, passes through the edges, and ends at the starting node.}
 present in it.
In every iteration, there are two steps. In Step 1, a belief message is passed from an FN to a VN, and in Step 2, the message is passed from a VN to an FN.  MPA starts by passing messages from leaf nodes.  A node has a degree (number of edges connected to it) $d$, then it will remain idle until  messages have arrived on $d-1$ edges.

Let $\eta_{j \rightarrow k}$  be the message sent from the  VN $j$ to the  FN $k$ and $\eta_{k \rightarrow j}$ be the message from the FN $k$ to VN $j$. Let $\zeta_j$ and $\xi_k$ be the set of nodes directly connected to  VN $j$ and  FN $k$,  respectively. %Let $A_j$ be the set of information symbols  that can be sent from VN $j$ and $a_j \in A_j$.
Next, MPA can be carried out as follows: 
\begin{enumerate}
\item 	Passing message from FN $k$ to VN $j$:
\begin{equation}\label{fn_vn_spa}
\begin{split}
    \eta_{k \rightarrow j}(\textbf{x}_j^m) &=  \sum_{{\sim}\textbf{x}_j} \left(\psi_k \prod_{j' \in \xi_k \setminus \{j\}} \eta_{j' \rightarrow k}(\textbf{x}_{j'}^m)\right), %\\     &~~~~~~~~~~~~~~~~~~~~~~~~~~~~~~~\text{for} ~a_j \in A_j,
    \end{split}
\end{equation}
where $\psi_k$ is the likelihood function associated with FN $k$ and $j' \in \xi_k \setminus \{j\}$ denotes all the VNs of $\xi_k$ except $j$th VN. Also, $\eta_{j' \rightarrow k}(\textbf{x}_{j'}^m)$ denotes the message from the VN $j'$ to the FN $k$ corresponding to symbol $\textbf{x}_{j'}^m \in \textbf{CB}_j, m=1,\cdots,M$.
\item	Passing message from VN $j$ to FN $k$:
\begin{align}\label{vn_fn_spa}
    \eta_{j \rightarrow k}(\textbf{x}_j^m) =\prod_{k' \in \zeta_j \setminus \{k\}} \eta_{k' \rightarrow j}(\textbf{x}_j^m),%~~~ \text{for} ~a_j \in A_j.
\end{align}
where $k' \in \zeta_j \setminus \{k\}$ denotes the FNs in $\zeta_j$ except the $k$th FN. The message is normalized to ensure that the sum of all the probabilities is equal to one.
\begin{align}\label{vn_fn_spaa}%\nonumber
    \eta_{j \rightarrow k}(\textbf{x}_j^m) =\frac{\prod_{k' \in \zeta_j \setminus \{k\}} \eta_{k' \rightarrow j}(\textbf{x}_j^m)}{\sum_{\textbf{x}_j^l \in \textbf{CB}_j} \prod_{k' \in \zeta_j \setminus \{k\}} \eta_{k' \rightarrow j}(\textbf{x}_j^l)  }.% \\%\hspace{.5cm}%    \text{for} ~a_j \in A_j.
\end{align}

\end{enumerate}
The above algorithm  uses the sum and product operations, and hence it is also called the sum-product algorithm. The detailed explanation of (\ref{fn_vn_spa}) and (\ref{vn_fn_spa}) is given in \cite{facsum}. It is noted that  (\ref{vn_fn_spa}) is simpler than (\ref{fn_vn_spa}) because there is no local function associated with a VN. After a few iterations, each bit's bit log-likelihood ratio (LLR) is calculated to estimate the bits transmitted by each user.
The MPA for decoding of $K \times J$ SCMA system is summarized in \textbf{Algorithm \ref{mpa_algo1}}.

\begin{algorithm}[!ht]
\caption{The Message Passing  Algorithm}
\label{mpa_algo1}
\footnotesize
\textbf{Inputs}: \\
$J$ : Number of users.\\
$K$ : Number of REs.\\ 
$\textbf{CB}_j$ : CB of $j$th user, $\forall j=1,\cdots,J$.\\ 
$\textbf{y}$ : The received signal vector.\\
\textbf{Output}:\\
Estimated value of bits for $j$th user,
  $\forall j=1,\cdots,J$.\\
\textbf{Initialize}:\\{ The prior probability of each codeword: $\eta_{j \rightarrow k}^0= P(\textbf{x}_j^m)= \frac{1}{M}, \forall j=1,\cdots,J, ~ k \in \zeta_j $.
 }
\label{algo:on-grid2}
\begin{algorithmic}[1]% enter the algorithmic environment
\STATE {\textbf{Step 1: Message passing along the nodes}} \\
  
  \For{$t =1~\KwTo~N_{t}$}
{a) Update Resource Node:
\begin{align*}
    &\eta_{k \rightarrow j}^t(\textbf{x}_j^m) =\\& \sum_{{\sim}\textbf{x}_j} \Bigg( \exp \left\{\frac{-1}{N_0} \norm{y_k- \sum_{j \in \xi_k} h_{k,j}x_{k,j}^m}^2 \right\} \\ & \prod_{j' \in \xi_k \setminus \{j\}} \eta_{j' \rightarrow k}^{t-1}(\textbf{x}_{j'}^m)\Bigg),\\
    \end{align*}
    
    b) Update Variable Node:
\begin{align*}
    \eta_{j \rightarrow k}(\textbf{x}_j^m) = P(\textbf{x}_j) \prod_{k' \in \zeta_j \setminus \{k\}} \eta_{k' \rightarrow j}^{t-1}(\textbf{x}_j^m).
\end{align*} \\
\textbf{end} } 

\STATE {\textbf{Step 2: Bit LLR}\\
  The final belief computed at each VN is:}
  \begin{align}\label{user_belief}
     {I}(\textbf{x}_j^m)= \left(\frac{1}{M}\right)\prod_{k \in \zeta_j } \eta_{k \rightarrow j}(\textbf{x}_j^m)
  \end{align}
%\STATE{\textbf{Step 4:}}
{{ %Let the number of bits per symbol be $b=\text{log}_2(M)$.
The $i$-th bit LLR $(1 \leq i \leq \text{log}_2(M))$ at $j$-th VN is }
\begin{equation} 
\begin{split}
 \text{\text{LLR}}(b_j^i) & =\log \frac{P(b_j^i=+1)}{P (b_j^i=-1)}\,, \\
                    & =\log \frac{\sum_{\textbf{x}_j^m \in \textbf{CB}_j \vert b_j^i=+1} {I}(\textbf{x}_j^m)} {\sum_{\textbf{x}_j^m \in \textbf{CB}_j \vert b_j^i=-1} {I}(\textbf{x}_j^m)}\,.
\end{split}
\end{equation} 
  where, $\hat{b}_j^i=+1$ if $\text{LLR} (b_j^i)$ is positive otherwise $\hat{b}_j^i=-1$.
  }

\end{algorithmic}
\end{algorithm}

\subsection{Illustration of MPA Decoding for $4 \times 6$ SCMA System}

 MPA is applied over the factor graph shown in Fig. \ref{fig:scma_fac} and messages are passed between the VNs and FNs in both the directions of the SCMA factor graph.  It is noted that  the factor graph of the SCMA system contains cycles in it. Thus messages are passed between the nodes for some iterations until the termination criteria is achieved. The objective is to detect the symbols transmitted by each user, i.e., compute (\ref{mpf}) using MPA, and this  can be carried out in the following steps:
\begin{enumerate}
    \item Initialization.
    \item Passing of messages between FNs and VNs.
    \item Termination and selection of codewords.
\end{enumerate}
The algorithm is discussed in a detailed manner for $4 \times 6$ SCMA system as follows:
\begin{itemize}
 \item Step 1: \textbf{Initialization}

 Assuming  received vector \textbf{y} and channel coefficient matrix \textbf{H} are known at receiver, firstly the likelihood ratio %$f(y_k \vert \textbf{X})$% 
 is computed at each FN. Let us assume a FN $l$ where data of three users corresponding to set $\xi_l=\{v_1.v_2,v_3\}$ superimpose. Therefore, the likelihood function at FN $l$ becomes $\psi(y_l \vert \textbf{x}_1,\textbf{x}_2, \textbf{x}_3,N_0)$, where,  $\textbf{x}_1,\textbf{x}_2,\textbf{x}_3$ are the codewords transmitted by users belonging to set $\xi_l$,  respectively. Assume the set of codeword elements allotted to user $v$ on RE $l$ is denoted as $\textbf{CB}_{l,v}$. The likelihood function of FN $l$ is given as
\begin{align}\nonumber
        \psi(y_l \vert \textbf{x}_1,&\textbf{x}_2,\textbf{x}_3,N_0 )= \exp \bigg(\frac{-1}{N_0} ||y_l-(h_{l1} x_{l,1} \\ \nonumber 
       &+h_{l2} x_{l,2} +h_{l3} x_{l,3} ||^{2} \bigg), \\ & \text{for}~ {x}_{l,1} \in \textbf{CB}_{l,1}, {x}_{l,2} \in \textbf{CB}_{l,2}, {x}_{l,3} \in \textbf{CB}_{l,3}.
\end{align}
Here, $x_{l,v}$ denotes the codeword element transmitted by the $v$th  user on the $l$th RE. In total, $K{M}^{d_f}$ values %\footnote{The likelihood ratio needs to be calculated for all combinations of symbols that user can send to neighboring RE.}
are stored for function $f(y_l \vert \textbf{x}_1,\textbf{x}_2,\textbf{x}_3,N_0 )$.  For an uncoded SCMA system, let us assume  equal prior probability for each codeword, i.e., $P(\textbf{x}_1 )$=$P(\textbf{x}_2 )$=$P(\textbf{x}_3 )$=$ \frac{1}{M}$. Therefore, the  initial message passed from  VN $v_1,v_2,v_3$  to the $l${th} FN is 
 \begin{align}
     \eta_{v_1 \rightarrow l}^{\text{init}} (\textbf{x}_1 )=\eta_{v_2 \rightarrow l} ^{\text{init}} (\textbf{x}_2 )=\eta_{v_3 \rightarrow l}^{\text{init}} (\textbf{x}_3 )=\frac{1}{M}\,.
 \end{align}

\item Step 2: \textbf{Passing of messages between FNs and VNs.}

%\begin{itemize}
    
     	\textbf{(i) Message from FN to VN}:\\
Let us assume  $\xi_l=\{v_1,v_2,v_3\}$, where $v_1,v_2,v_3$ denotes the three users connected to RE $l$, respectively.  To pass the message  from the FN to one user, information received on the FN from the other two users may be regarded as extrinsic information, as shown in Fig. \ref{fig:scma_fn_vn}. 
    %   trim=left bottom right top
%\fbox{\includegraphics[scale=0.5,trim=3.75cm 6cm 4.5cm 6cm,clip]{fn_to_vn_scma.pdf}}

\begin{figure}
\centering
\includegraphics[scale=0.3,trim=3.75cm 6cm 4.5cm 6cm,clip]{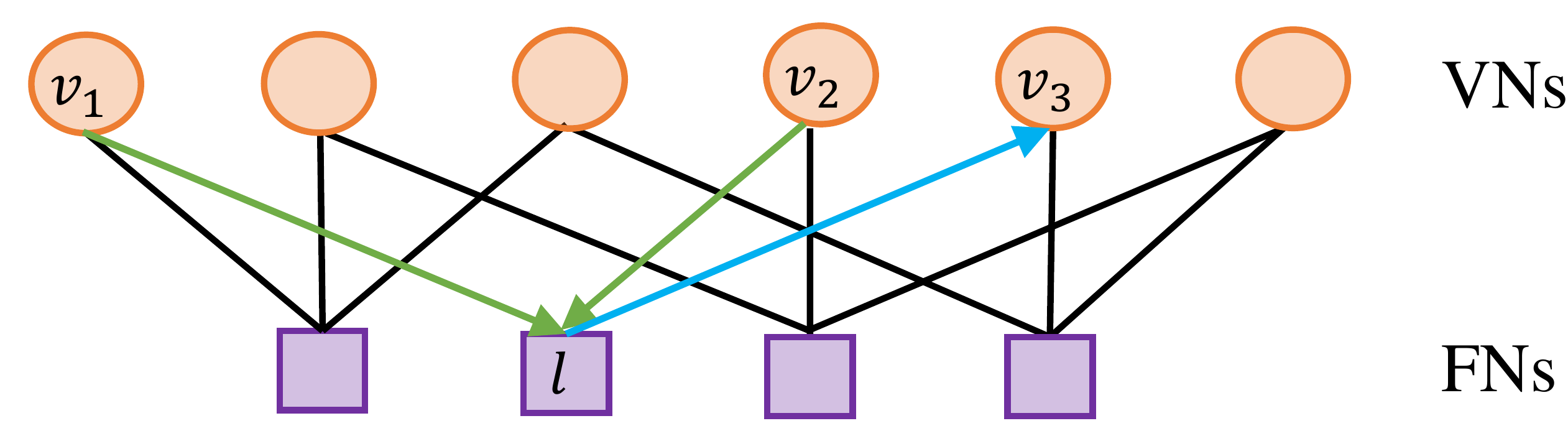}
\caption{Message Passing from FN to VN.}
\label{fig:scma_fn_vn}
\end{figure}
    
Message passed from FN $l$ to VN $v_1$ is given as
\begin{multline}\label{Fn_vn_spa}
   % \begin{split}
        \eta_{l \rightarrow v_1} (\textbf{x}_1)\\= \sum_{\textbf{x}_2\in \textbf{CB}_2} \sum_{\textbf{x}_3\in \textbf{CB}_3} \bigg(\psi(y_l \vert \textbf{x}_1,\textbf{x}_2,\textbf{x}_3,N_0 )\\  \times \eta_{v_2 \rightarrow l} (\textbf{x}_2 ) \hspace{0.05cm} \eta_{v_3 \rightarrow l} (\textbf{x}_3)\bigg) \hspace{0.25cm} \text{for} ~ \textbf{x}_1 \in \textbf{CB}_1.
      %  \end{split}
\end{multline}

In (\ref{Fn_vn_spa}), messages from the two VNs (i.e., $\eta_{v_2 \rightarrow l} (\textbf{x}_2 )$ and $\eta_{v_3 \rightarrow l} (\textbf{x}_3 )$) are multiplied with the local likelihood  function of $l$th FN and then marginalized with respect to $v_1$. Similarly, message passed from FN $l$ to VNs $v_2$ and $v_3$, respectively are 

\begin{multline}
    %\begin{split}
    \eta_{l \rightarrow v_2} (\textbf{x}_2 )\\= \sum_{\textbf{x}_1\in \textbf{CB}_1} \sum_{\textbf{x}_3\in \textbf{CB}_3} \bigg(\psi(y_l \vert \textbf{x}_1,\textbf{x}_2,\textbf{x}_3,N_0 )\\ 
\times      \eta_{v_1 \rightarrow l} (\textbf{x}_1 ) \hspace{0.05cm} \eta_{v_3 \rightarrow l} (\textbf{x}_3)\bigg) ~~ \text{for} ~ \textbf{x}_2 \in \textbf{CB}_2. 
   % \end{split}
\end{multline}

\begin{multline}
   % \begin{split}
    \eta_{l \rightarrow v_3} (\textbf{x}_3 )\\= \sum_{\textbf{x}_1\in \textbf{CB}_1} \sum_{\textbf{x}_2\in \textbf{CB}_2} \bigg(\psi(y_l \vert \textbf{x}_1,\textbf{x}_2,\textbf{x}_3,N_0 )\\ 
\times     \eta_{v_1 \rightarrow l} (\textbf{x}_1 )\hspace{0.05cm}  \eta_{v_2 \rightarrow l} (\textbf{x}_2)\bigg)\, ~~ \text{for} ~\textbf{x}_3 \in {\textbf{CB}_3}.
%\end{split}
\end{multline}
Message passing from FN $l$ to VN $v_3$ is shown graphically in Fig. \ref{fig:scma_fn_vn}. The message transmitted from FN $l$ to VN $v$ indicates the guess of what signal is received at FN $l$ for all possible values of VN $v$.\\

  \textbf{ (ii) Message from VN to  FN}:\\
 Let us assume $\zeta_v=\{l_1,l_2\}$, where $l_1$ and $l_2$ are the FNs connected to VN $v$.\\
%   trim=left bottom right top
%\fbox{\includegraphics[scale=0.5,trim=3.75cm 6cm 4.5cm 6cm,clip]{vn_to_fn_scma.pdf}}

 \begin{figure}[h!]
  \centering
\includegraphics[scale=0.3,trim=3.75cm 6cm 4.5cm 6cm,clip]{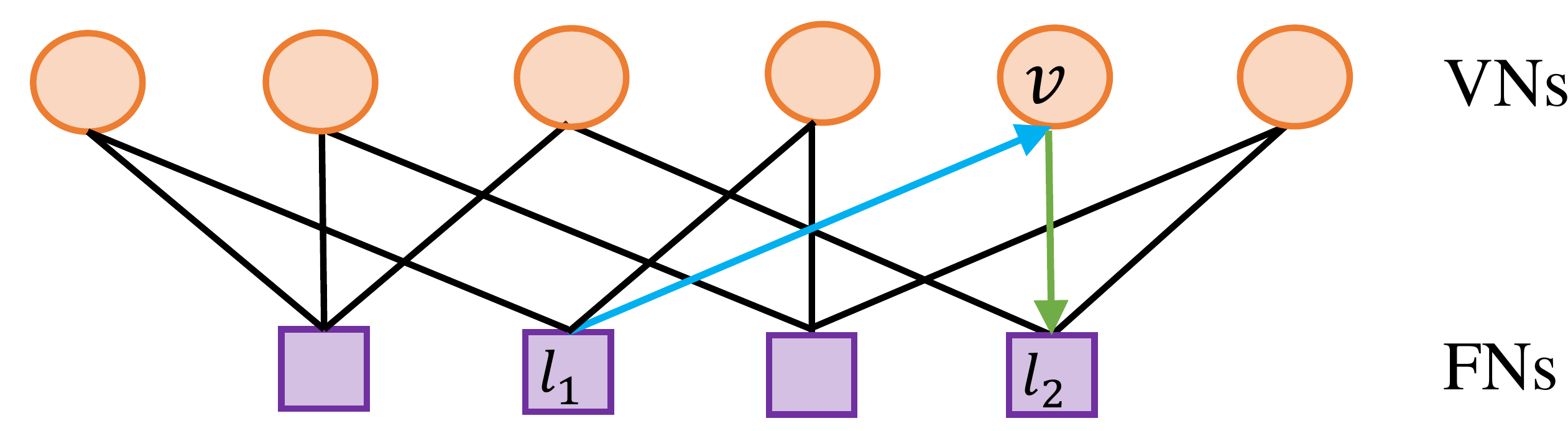}
\caption{Message Passing from VN to FN.}
\label{fig:scma_vn_fn}
\end{figure}
Message sent from VN $v$ to FN $l_1$ and $l_2$, respectively are:

\begin{multline}\label{vn_fn_1}
    \eta_{v \rightarrow l_1}(\textbf{x}_v)= \text{normalize}(P_a (\textbf{x}_v) \hspace{0.1cm} \eta_{l_2 \rightarrow v }(\textbf{x}_v)) \\
    \text{for}~ \textbf{x}_v \in \textbf{CB}_v ,
\end{multline}

\begin{multline}\label{vn_fn_2}
            \eta_{v \rightarrow l_2}(\textbf{x}_v)= \text{normalize}(P_a (\textbf{x}_v) \hspace{0.1cm} \eta_{l_1 \rightarrow v} (\textbf{x}_v))\, \\
            \hspace{1cm} \text{for}~ \textbf{x}_v \in \textbf{CB}_v,
\end{multline}
where $P_a$ denotes the  prior probability of user $v$ and $\eta_{l_2 \rightarrow v}$ indicates the updates VN $v$ received from the  FN $l_2$.
Here, normalization is necessary  to ensure  that each belief falls in the range of [0,1].
The normalization is conducted similar to  (\ref{vn_fn_spaa}) and so,  (\ref{vn_fn_1}) can also be written as
\begin{align}\nonumber
    \eta_{v \rightarrow l_1}(\textbf{x}_v) =\frac{ P_a(\textbf{x}_v) \eta_{l_2 \rightarrow v}(\textbf{x}_v)}{\sum_{\textbf{x}_v}  \eta_{l_2 \rightarrow v}(\textbf{x}_v)  } \\ \text{for} ~\textbf{x}_v \in \textbf{CB}_v.
\end{align}
The message from the VN $v$ to FN $l_2$ is shown graphically in Fig. \ref{fig:scma_vn_fn}.
%\end{itemize}
Since the factor graph has cycles, the passing of messages between FNs to VNs is repeated for a few iterations. This continues until no considerable change is observed in the belief computed at each VN. 

 \item   Step 3:\textbf{ Termination and selection of codewords.}

After repeating Step 2 for  $N_{\text{t}}$ number of iterations, the final belief is  computed at each VN, which is the product of the prior probability and messages from the neighboring FNs of each VN.
\begin{multline}    \label{term}
    	I_v(\textbf{x}_v)= P_a(\textbf{x}_v) \hspace{0.05cm}  \eta_{l_1 \rightarrow v} (\textbf{x}_v) \hspace{0.05cm} \eta_{l_2 \rightarrow v} (\textbf{x}_v)\,,  \\
    	\text{for}~ \textbf{x}_v \in \textbf{CB}_v.
\end{multline}
The probability for each codeword is calculated at each VN, and the codeword  with the highest probability becomes the estimated codeword for that user. This is one method of estimating the transmitted symbol for each user. Another method is to compute bit LLR from $I_v(\textbf{x}_v)$. The $i$-th bit LLR for VN $v$ is given as
\begin{equation} \label{bit_llr}
\begin{split}
 \text{LLR} ({b_v^i}) & =\log \frac{P(b_v^i=+1)}{P (b_v^i=-1)}\\\
                    & =\log \frac{\sum_{\textbf{x}_v \vert b_v^i=+1} I_v(\textbf{x}_v)} {\sum_{\textbf{x}_v \vert b_v^i=-1} I_v(\textbf{x}_v)}\,,\\ & \hspace{3cm} \text{for}~ i=1,\cdots,B.
\end{split}
\end{equation}
%   trim=left bottom right top
%\fbox{\includegraphics[scale=0.5,trim=10cm 4cm 12cm 7cm,clip]{llr_bit.pdf}}
\begin{comment}

\begin{figure}
\centering
\includegraphics[scale=0.5,trim=10cm 4cm 12cm 7cm,clip]{llr_bit.pdf}
\caption{Bit by bit decoding.}
\label{fig:llr_bitbit}
\end{figure}
\end{comment}

In case of $M=4$, there are two bits per symbol. Let $\textbf{CB}_v=\{\textbf{x}_v^1,\textbf{x}_v^2,\textbf{x}_v^3,\textbf{x}_v^4\}$ denote the four codewords corresponding to four symbols $\{s^1,s^2,s^3,s^4\}$ that can be sent by user $v$. Using  (\ref{term}), final belief corresponding to each of the four codewords is calculated. %It is to be noted that 4 codewords, i.e. $\textbf{x}_1,\textbf{x}_2,\textbf{x}_3,\textbf{x}_4$ corresponds to the 4 symbols $s_1,s_2,s_3,s_4$ that can be sent by user $v$. 
Then,  the LLR of the first bit is given as
\begin{equation} 
\begin{split}
 \text{LLR} (b_v^1) &=\log ⁡\frac{P (b_v^1=+1)}{P (b_v^1=-1)}\
                       =\log ⁡\frac{ I_v(\textbf{x}_v^1 )+I_v(\textbf{x}_v^2)}{I_v(\textbf{x}_v^3)+I_v(\textbf{x}_v^4)}\,.
\end{split}
\end{equation}
%It can be seen from Fig. \ref{fig:llr_bitbit} that first bit ($b_1$) is 0 for symbols $s^1$ and $s^2$ and is 1 for symbols $s^3$ and $s^4$, respectively. Thus, 
The ratio of the belief computed corresponding to codewords $\textbf{x}_v^1,\textbf{x}_v^2$ and $\textbf{x}_v^3, \textbf{x}_v^4$ gives the bit LLR of the first bit. Consequently, if $P(b_v^1=-1)>P(b_v^1=+1)$ or if $\text{LLR}(b_v^1)<0$, then `1' is decoded for first bit of the symbol otherwise 0. Similarly, for the second bit, if $\text{LLR} (b_v^2)<0$, then `1' is decoded for the second bit of the symbol. The bit LLR for the second bit is given as
\begin{align}
    \text{LLR} (b_v^2) = \log ⁡\frac{ I_v(\textbf{x}_v^1 )+I_v(\textbf{x}_v^3)}{I_v(\textbf{x}_v^2)+I_v(\textbf{x}_v^4)}\,.
\end{align}
\end{itemize}

 \begin{table}[]
 
 \begin{center}
 \caption{List of Notations}
 \begin{tabular}{||c | c  ||} 
 \hline
Parameter & Notation   \\ 
 \hline \hline
  Number of users & $J$ \\
  Number of resource elements & $K$ \\
   Transmitted codeword of the $j$th user & $\textbf{x}_j$
 \\
   Constellation size  & $ M$
\\
   Number of bits per symbol & $B$
\\
Column weight of the factor graph matrix & $d_v$ \\
Row weight of the factor graph matrix & $d_f$ \\
Set of users transmitting data over the $k$th RE & $\xi_k$\\
Set of REs connected to $j$th VN   & $\zeta_j$\\
Message passed from $k$th FN to $j$th VN & $\eta_{k \rightarrow j}$ \\
Message passed from $j$th VN to  $k$th FN & $\eta_{j \rightarrow k}$ \\
%Set of codewords allotted to the $j$th user & $ \textbf{CB}_j$\\
\hline
\end{tabular}
%\label{tab:kysymys}
\end{center}
 \end{table}

\section{Advanced SCMA Decoding Techniques}
In SCMA, MPA is an efficient decoding technique by exploiting the CBs sparsity.  Using MPA, the receiver is able to attain near-optimum decoding performance  with computation complexity  $O(d_f M^{d_f})$.
%The computational complexity of the conventional MPA detection algorithm increases exponentially with $d_f$. 
In this section, we first introduce several advanced MPA based detectors with lower computation complexity, followed by a few non-MPA based SCMA detectors. %In this section, we will discuss the MPA based detection methods.

\subsection{Discretized MPA (DMPA)}
%In each iteration of MPA, 
%In \cite{scmapot_140}, \cite{scmapot_141}, a detection algorithm based on discretization and fast Fourier transform (FFT) was proposed for SCMA system. 
In standard MPA, the dominant complexity is incurred by updating the messages of FNs (i.e., Step 1 (a) of \textbf{Algorithm 1}), where the message from an FN
to a VN is computed based on the messages coming
from the other $d_f- 1$ VNs  connected to the respective FN.  In \cite{scmapot_140,scmapot_141}, the update of messages at the FN is investigated from
a new perspective. In MPA, each VN is treated as a node that can take $M$ values, but in DMPA, each VN is considered an RV on a complex field. The PDF $f_j(t)$\footnote{If $f(t)$
is a function, let $f$ denote its sampling sequence and the $f[n]$
is the $n$th element of the sequence.} of the $j$th VN is all zero except $M$ impulses.% eqn 6 of paper.  

%The PDFs of the $d_f- 1$ RVs are discretized and convolved to get the new message. 

%In DMPA, the main idea is to discretize the PDF of each VN, which means to shift the impulses of each PDF to nearest sampling points. 
In this work, \emph{Step 1 (a) of \textbf{Algorithm 1} is proved to be  equivalent to the convolution of $d_f$ PDFs}
at the point of one codeword element.
Assume that  the constraint to update FN is $x_1+ x_2+\hdots +x_{d_f}+n_o=y$, where the corresponding PDFs are $f_1(t), f_2(t),\hdots, f_{d_f}(t), \mathcal{N}_p(t)$, respectively. %The PDF of $f_1(t)$ consists of only M impulses and $(f_2 * \cdots *f_{d_f})(t)$ consists of $M^{d_f-1}$ impulses, respectively. 
Let $g(t)= (f_2 * \cdots *f_{d_f})(t) * \mathcal{N}_p(t) $, then it is proven that % and is computed as the superposition of $M^{d_f-1}$ noise PDFs. When $f_1(x)$ is updated using the constraints, it can be expressed as 
%\begin{align*}
 %   f_1(x_{k1m})=g(y_k-x_{k1m})
%\end{align*}
%This means that the FN update step in MPA is basically computation of convolution. Thus Equation (\ref{fn_vn_spa}) can be given as 
\begin{align}\label{dmpa_n}
    \eta_{k \rightarrow j}(\textbf{x}_j^m) = g(y_k-x_{k,j}^m).
\end{align}
%where $x_{k,j}^m$ is the codeword element sent by the $j$th user on the $k$th RE for the $m$th symbol.
The above \emph{convolution operation can be efficiently computed by discretization and FFT}. %Let $f$ denote the sampling sequence and $f[n]$ denote the $n$th sampling point of the function $f(t)$, respectively. 
%Again assuming the constraint $x_1+ x_2+\hdots +x_{d_f}+n_o=y$.
Suppose all the possible values of $x_j (j \in \xi_k)$
are  in the interval $[-wid;wid]$. Since, the PDF of noise is continuous and has integral one, there exists an $n_o$ such that $\eta(t)$ is negligible
outside $[-n_o ;n_o]$. Let $w$ denote the precision of discretization and $n = \{0, 1,\cdots, (2wid)/w\}$. 
Then, the integral of each impulse is distributed to the nearest sampling point.

%The discretized version of noise PDF $\mathcal{N}_p[n]$ is generated by directly sampling $\mathcal{N}_p(t)$ with sampling interval $w$. 
Using the circular convolution theorem, the sampling sequence of $g(t)$ can be computed with the help of FFT as follows
\begin{multline}\label{dmpa_g}
    g=\mathcal{F}^{-1}[\mathcal{F}(pad(f_2))\bullet \mathcal{F}(pad(f_3))\bullet \cdots \bullet \mathcal{F}(pad(f_{d_f}))\\ ~~~~~~~\bullet \mathcal{F}(pad(\mathcal{N}_p)],
\end{multline}
where $\mathcal{F}$ denotes the FFT, $\bullet$ denotes the Hadamard product and $pad()$ pads zeros to the end of the sequence, respectively. Thus, the FN update process can be summarized as follows:
\begin{enumerate}[label=(\arabic*)]
    \item For every connected edge $(j,k)$, the discretized version of $\eta(t)$  and  $\{f_{j}(t)\}$ for $m=1,\hdots, d_f -1$ are computed.% based on $\eta_{j' \rightarrow k}^{t-1}$, where $j'_m \in \xi_k \setminus\{j\} $.
    
    \item The discretized sequences are padded with zeros until they have length $N$, where $N$ is the size of  FFT.
    
    \item Compute $g$ from (\ref{dmpa_g}).
    \item The FN is updated using (\ref{dmpa_n}).
\end{enumerate}
With the above modifications,  the complexity
to update one message becomes the complexity of the 2-D FFT algorithm, which
is $O(d_f^2 \text{ln}(d_f ))$  and the number of operations increases linearly with $M$. Thus in DMPA, the FN update step is proved to be equivalent to computing
a convolution. Then, the convolution is solved by
discretization and FFT. Regarding the detection accuracy, as the sampling interval decreases, the performance of DMPA and MPA comes closer, and both have negligible error performance difference for the sampling interval $w=0.05$. 

%\subsection{Improved Log-MPA}
\subsection{Sphere Decoded MPA}
To simplify the conventional MPA, \cite{scmapot_110} proposed a logarithm domain MPA (Log-MPA) detector, which reduced the product operations involved without compromising  the performance. To further simplify the detection, \cite{scmapot_108,scmapot_109} approximated the sum of exponential operations as a maximum operation, and it is called Max-Log-MPA. This reduces complexity but with slight performance degradation. Another way of reducing the MPA complexity is to restrict the search space  around the received signal vector, and this is known as sphere decoded MPA (SD-MPA) \cite{scmapot_144,scmapot_145,scmapot_150,scmapot_151}.

An improved  Log-MPA  Region-Restricted detector (RRL detector) was proposed in \cite{scmapot_144} for SCMA downlink systems. %The FN update step in standard MPA (i.e., Step 1 (a) of \textbf{Algorithm 1}) contributes the most for increasing the MPA complexity.
In the case of a downlink system, the received signal at the $k$th RE is given as:
\begin{align}
    y_k= h_k \sum_{j \in \xi_k} x_{k,j} + n_k \triangleq h_k z_k +n_k,
\end{align}
where $z_k = \sum_{j \in \xi_k} x_{k,j} \in Z_k$, the constellation $Z_k$ is called superposition constellation and $|Z_k| = M^{d_f}$ which increases exponentially with $d_f$. In the RRL detector, the superposed constellation points are divided into several regions. Based on each region $\mathfrak{D}$, a search region $\mathcal{S}$ is obtained (as shown in Fig. 3 \cite{scmapot_144}). In RRL detector, the superposed constellation points lying in the search region  $\mathcal{S}$ are only involved for FN update. This reduces the unnecessary computations with respect to constellation points outside  $\mathcal{S}$. With this change, the FN update step is given as
% eqn 5 or 23 of paper
\begin{align}
    \eta_{k \rightarrow j} (\textbf{x}_{j} )= \underset{z_k \in \mathcal{S}: \textbf{x}_j} {{F}_m} \left(  \sum_{ j' \in \xi_k \setminus j} \eta_{j' \rightarrow k} (\textbf{x}_{j'} ) + \mathcal{M}(z_k)  \right)
\end{align}
% eqn 23, eqn 4 of RRL Paper.
where
\begin{align}\label{yhx_norm}
    \mathcal{M}(z_k)=  -\frac{1}{2 \sigma^2} |y_k-h_k z_k|^2,
\end{align}
% eqn 22
\begin{align}\nonumber
    {{F}_m} \triangleq \overset{\sim}{max}&(\mathsf{s}_{n'}, max(\mathsf{s}_{{n'}-1},\hdots, max(\mathsf{s}_2, \mathsf{s}_1))), \\
   & \text{for}~ n'=M^{d_f -1},\\\nonumber
   % eqn 15 ke baad
    \mathsf{s}_{n'}=& \sum_{ j' \in \xi_k \setminus j} \big(\mathcal{M}({z_k}) + \eta_{j' \rightarrow k} (\textbf{x}_{j'} ) \big) \\& %~~ \text{for} ~m' \in \Xi_{n'}, |\Xi_{n'}|=n' \\
    \text{and}, ~
    \overset{\sim}{max}(a',b') = max (a',b')+C_1 (a'-b'), 
\end{align}
% eqn 22, 6,7
    where  $C_1(x)=\text{ln} (1+e^{-x})$ is the correction function. %non-linear equation \cite{rrldetector_5}
    %\[
  %C_1(a')=  
%\begin{cases}
 % \frac{1.08}{1.16+a'}-0.19 & \text{a}'<5 \\
  %0 & \text{otherwise}
%\end{cases}
%\]
In this manner, the FNs are updated while   restricting the search space,  reducing the detection complexity.
The SD-MPA detectors proposed in
\cite{scmapot_144,scmapot_145,scmapot_150,smapot_151_ka_4} were designed for   constant modulus constellations (e.g., QPSK) or with a subset of non-constant modulus constellations (e.g., M-QAM). So, for generalized SCMA constellations, a new improved SD-MPA based detector was proposed in \cite{scmapot_151}, called GSD-MPA. \color{black}However, this method is sensitive to the overloading factor. Therefore, to further improve the performance with low  complexity, \cite{sgsd_mpa} proposed two pruning algorithms and introduced the simplified
GSD-SCMA (SGSD-SCMA). \color{black}%Let $\textbf{c}_L=[c_1,\cdots,c_L]^T$ be $L= d_v J$ dimensional vector, which contain the transmitted constellation points of $J$ users. Let $\textbf{c}_L=[\textbf{c}^1 \textbf{c}^2]^T$ where $\textbf{c}^1$ and $\textbf{c}^2$ denotes the first $K$ and $L-K$ elements of $\textbf{c}^L$, respectively. The Maximum Likelihood detection problem can be re-written as
%\begin{align}
 %   \hat{\textbf{x}}=  \mathop{\arg \min}_{\textbf{x}^{1} \in \chi^1}
%\end{align}

\subsection{Scheduled MPA}
In  \cite{scmapot_111}, MPA was simplified and optimized from two aspects. Firstly, a lookup table (LUT) was designed to approximate the correction function $C_1(x)$  to reduce the computation  complexity of $\overset{\sim}{max}$ operation  for the FN update. For this, the range of $C_1(x)$ function is divided into $\gamma$ quantization  intervals, which means LUT consists of $\gamma$ items.  The larger the number of intervals, the higher the approximation precision, but it comes with higher complexity. Next, the authors proposed two enhanced scheduling schemes for implementing MPA: single scheduling MPA (SS-MPA) and 
multiple scheduling MPA (MS-MPA).   In standard MPA, the FNs and VNs are updated using the information received in the previous iteration. In scheduled MPA, instead of updating all the FNs at one step and then updating the VNs at the second step, the FNs and VNs are jointly updated to speed up the convergence of MPA detection. For every VN $j$, if all the FNs  belonging to $\zeta_j $ are updated, then VN $j$ will be updated instead of waiting for all FNs to be updated.  This speeds up the update process. This scheme is called SS-MPA-VN since messages are updated and passed VN-by-VN. In this, the initialization step and bit LLR step are same as standard MPA. In this, the FN update step  of the standard Log-MPA gets modified as follows:
% eqn 28
\begin{align}\nonumber \label{fn_ssmpa}
    \eta_{k \rightarrow j}(\textbf{x}_j) &=  \overset{\sim}{max}_{\sim \textbf{x}_j} \bigg(\psi_k + \sum\limits_{\substack{j'\in \xi_k \setminus \{j\} \\ {\pi}^{-1}(j') < v' }} \eta_{j' \rightarrow k}^{t}(\textbf{x}_j) + \\
   & \sum\limits_{\substack{j'\in \xi_k \setminus \{j\} \\ {\pi}^{-1}(j') > v' }} \eta_{j' \rightarrow k}^{t-1}(\textbf{x}_j)\bigg), ~~\text{for} ~\textbf{x}_j \in \textbf{CB}_j,
\end{align}
where function $j = \pi (v')$ denotes the VN update order, where $v' = 1, 2,\cdots , J$ and $j$ 
 denotes the updating VN index, ${\pi}^{-1}(\cdot)$ denote the inverse function of $\pi (\cdot)$.
%where $v=1 \rightarrow 2 \rightarrow \cdots \rightarrow J$ denotes the VN order.
A similar procedure can be done with respect to FN too, where each FN $k$ will be updated if $\xi_k$ are updated. That scheme will be called SS-MPA-FN. To further speed up the detection method, MS-MPA can be used where a number of SS-MPA based detectors work in parallel with different update orders. %SS-MPA at each VN  exchange their most reliable messages with each other. 

\subsection{PM MPA}
The partial marginalization based MPA (PM-MPA) was introduced in \cite{scmapot_121_35}. 
% padha humne 121_35 vala h.
The FNs and VNs are updated as in conventional MPA for $t'$ number of iterations. After $t'$ number of iterations, the PM-MPA computes the codeword for the last $u'$ number of users. For the rest $N_t-t'$ iterations, the message for the remaining $J-u'$ users are updated with the already determined codewords.  %With $N_t$ iterations, the codewords for $J-v'$ users are determined.
In \cite{scmapot_121_35}, $u'$ users are chosen 
randomly in $t'$th iteration, so  some unreliable codewords may get chosen, which deteriorate the BER performance.

To improve the BER performance with same complexity, \cite{scmapot_121} introduced Improved PM-MPA in which the $u'$ users are chosen using a parameter $\textsf{w}$. This parameter $\textsf{w}$  measures the reliability of codewords of
all the users.  After $t'$ iterations, the final belief $I_j^m$ is calculated for $1 \leq j \leq J, ~ m=1,\cdots, M$ (as shown in (\ref{user_belief})). Then these beliefs are sorted as
\begin{align*}
    \textbf{L}_j= sort (\{I_j^1,\cdots,I_j^M\}, ~\text{`descend'}).
\end{align*}
The reliability of all codewords of all users is calculated.
\begin{align*}
    \textsf{w}_j= \frac{\textbf{L}_j [1]}{\textbf{L}_j[2]}
\end{align*}
The $u'$ users are chosen based on their reliability, and their codewords are determined. In the rest of the iterations, messages of undetermined users get updated with the codewords of determined users. The complexity of PM-MPA can get reduced to 50\% of the complexity of standard MPA for $M=4$ depending on the user parameters values.

\subsection{Extended Max-Log MPA}
In \cite{scmapot_122}, authors proposed extended max-log (EML) MPA, which jointly reduces the complexity for VNs and FNs update at the SCMA receiver. Due to the sparsity of $\textbf{F}_{K \times J}$, SCMA can be
represented by the Tanner graph as well. Non-zero elements in $\textbf{F}_{K \times J}$
represent the connection between FNs and VNs. The messages passing over edges on the Tanner graph are $q$ dimension
vectors. A $q$-order Galois field
 contains  elements
$(\{0, 1, \cdots, q-1\} \in GF(q))$. The SCMA codewords are mapped to the $q$-order Galois field as
\begin{align}\label{emlmpa_1}
    f' : \textbf{A}_{MC} \rightarrow GF(M).
\end{align}
This indicates that each constellation point of $\textbf{A}_{MC}$ is mapped to an element of $GF(M)$. A trellis diagram can represent the FNs update in SCMA according to  (\ref{emlmpa_1}). The vertices in the trellis diagram correspond to the complex symbols of the CB, and the connecting edges represent the possible paths that need to be searched  for calculating $\mathcal{M}(z_k), ~ \forall ~ m \in M$ (as shown in (\ref{yhx_norm})). All the paths of this trellis diagram form a combination set $\Xi$ and $\vert \Xi \vert = M^{d_f}$. 

%After computing the messages at each VN, most reliable $m_c (m_c < M)$ symbols are chosen at the VNs after one iteration. These symbols are chosen based on a reliability parameter and once $m_c$ is chosen, it will be fixed in this algorithm. Thus, 
The main difference between MPA and EML-MPA is the initialization and FNs update. These two steps are as follows:

\begin{itemize}
    \item Initialize: The reliability of each constellation point at VNs is
calculated as:
    \begin{align}
        RL_j(m)= \sum_{d=1}^{d_v} U_{dj} (m), ~\text{for}~ m=1,\cdots,M,% below statement eqn(12) scmapot_122
    \end{align}
    where $U_{dj}$ is the maximum of the logarithm of $\eta_{d\rightarrow j}$ over the possible constellation points.
    The parameter $RL_j(m)$  denotes the reliability  for each symbol.
    Based on $RL_j$, $m_c$ constellation points are chosen and initial messages are passed from these $m_c$ number of nodes. 
    
    \item FNs update: FNs are updated for only $m_c$ constellation points with $ \norm{y_k- \sum_{j \in \xi_k} h_{k,j}x_{k,j}^m}^2, ~ \forall ~ m \in m_c$.
\end{itemize}
It is to be noted that only $m_c < M$ dimension messages are updated in the above two steps. Consequently,  the computational complexity for  VNs update is also reduced. The standard MPA has $M^{d_f}$ search paths,  which now reduces to  $m_c^{d_f}$ in EML-MPA, and thus, overall receiver complexity decreases.

In \cite{scmapot_122}, the truncation parameter $m_c$ is fixed, which makes the BER performance loss for EML-MPA. To strike a balance between BER and complexity, \cite{scmapot_123} proposed a dynamic trellis based MPA (DT-MPA).  In DT-MPA, the parameter $m_c$ decreases as the number of iterations increases. Also, $m_c$ varies for each user as different users have different channel conditions. For this, a threshold based decision method was introduced. %The trellis diagram of $k$th FN contains $d_f$ columns and $M$ rows. Firstly, the message values at each column node is sorted in descending order, and then take the difference of the message values, i.e.
%\begin{align}
 %   r_{j,i}^t= Q_{j,sort}^{t-1}(x_i)- Q_{j,sort}^{t-1}(x_{i+1}), ~~ i=1,\cdots, m_{cj}^{t-1}-1
%\end{align}
%where $Q_{j,sort}^{t-1}$ represents the sorted message vector of $j$th user after $t-1$ iterations, $x_i$ denotes the $i$th symbol from the sorted message and $m_{cj}^{t-1}$ denotes the $m_c$ of $j$th VN after $t-1$ iterations. Next, the difference vector $\textbf{r}_j^t= [r_{j,1}^t, \cdots, r_{j,m_{cj}^{t-1}-1}^t]$ is generated. The elements of $\textbf{r}_j^t$ are compared with threshold $\tau$. If the $i$th element of $\textbf{r}_j^t$ is greater than $\tau$, then $m_c=i$ and $i$ symbols with large message values are left out. If $r_{j,i}^t < \tau $, then $\tau$ is compared with $i+1$th element and, so on. Thus, the parameter $m_{cj}$ is updated with each iteration and only FNs and VNs are updated for $m_{cj}$ symbol values. 
This reduction of $M$ from standard MPA to DT-MPA reduces complexity without compromising  the BER performance. 

\subsection{Detection of Coded SCMA System}
SCMA can  be combined with 
channel decoders to improve transmission reliability, such as
low-density parity-check (LDPC) codes, polar codes, and so
on \cite{scmapot_153,scmapot_154,scmapot_156,mmse_relted_papers_15,mmse_relted_papers_16}.

\subsubsection{LDPC Coded SCMA System}
%\subsection{MMSE-PIC}
%The wireless communication link requires error correction coding like polar codes and low-density parity-check (LDPC) codes as adopted by 5G standard \cite{scmapot_153,scmapot_154}. 
One popular approach for multi-user detection is to attempt the removal of multi-user interference from each user's received signal, among which parallel interference cancellation (PIC) is a promising candidate. 
In this, each user  attempts
to cancel its multi-user interference. As
compared with the serial processing scheme, since the interference cancellation
is performed in parallel for all users, the delay required to
complete the operation decreases significantly\cite{pic_paper}.

In literature, a variety of sub-optimum algorithms have been proposed in the literature, to compute the extrinsic LLRs\cite{asic_mmsepic_13, asic_mmsepic_15}.
In \cite{scmapot_156}, a hardware feasible minimum mean-square error (MMSE) PIC algorithm is proposed for LDPC coded SCMA detection. At the transmitter, the message bits of all users are coded by the LDPC encoder, followed by the SCMA encoder. 
%Let $\textbf{x}_j=[x_{1,j},\cdots,x_{K,j}]$ be the SCMA codeword of the $j$th user. 
Initially, the \emph{a priori} bit LLR values $L_{j}^{A,i}$ are initialized for
all VNs.
The MMSE-PIC based message passing detection is  discussed below:

%\emph{Operations at User node:}
%In the $t$-th SCMA iteration, $j$th user node receives  $i$-th bit LLR $LU^{t-1,i}_{k' \rightarrow j}$ $(1 \leq i \leq \text{log}_2(M))$ for $x_{k,j}$ from $k'$ resource node.  Then,  $i$-th bit LLR $LV^{t,i}_{j \rightarrow k}$  to $k$th FN is calculate as
%\begin{align}
 %   LV^{t,i}_{j \rightarrow k}= L_{S,j}^{A,i}+ \sum_{k' \in \zeta_j\setminus \{k\}} LU^{t-1,i}_{k' \rightarrow j}
%\end{align}
% eqn 18
% \eta symbol humre khyal se symbol LLR dega, and usse bit LLR nikalne ke loye use eqn8. 
%where $L_{S,j}^{A,i}$ denotes the  \emph{a priori } bit LLR for the SCMA detector, and $LU^{t-1,i}_{k' \rightarrow j}$ denotes the bit LLR from $k'$ FN at $(t-1)$ iteration. \\
%\emph{Operations at Resource node:}
\begin{itemize}
    \item \emph{Computation of Soft-Symbols}:
    The algorithm starts by computing estimates of the transmitted symbols according to \cite{asic_mmse_21}
\begin{align}\label{est_xkj}
    \hat{x}_{k,j}= \sum_{a \in \textbf{CB}_{k,j}} P(x_{k,j} =a) a
\end{align}
where %$x_{k, j}$ is a modulation symbol from a constellation $\mathbb{X}$ of size M, 
$P(x_{k,j} =a)= \prod_{i=1}^B \frac{1}{2} (1+ 2a_{i}-1) \text{tanh} (\frac{1}{2} LV^{t,i}_{j \rightarrow k})$ is the probability of modulation symbol $a$ and $a_i$ is the $i$-th bit
of symbol $a$. The reliability of each soft-symbol is characterized by its variance
\begin{align}
    \varepsilon_{k,j}^t= E [\vert x_{k,j} - \hat{x}_{k,j} \vert ^2]
\end{align}    

\item \emph{Parallel Interference Cancellation (PIC)}:  Let $\textbf{h}_{\textbf{r}k}=[h_{k,j}]$, where $j \in \xi_k$, denotes the channel vector for FN $k$. To detect the symbol transmitted by $j$th user, interference from other users can be cancelled using previously computed soft-symbols $\hat{x}_{j'}$ (\ref{est_xkj}), where $j' \neq j$. Let $\textbf{G}_{\textbf{r}k}= \textbf{h}_{\textbf{r}k}^\text{H} \textbf{h}_{\textbf{r}k}$, then each column of $\hat{Y}_{\textbf{r}k}$ can be given as
\begin{align}\label{y_npi}
\hat{y}_{\textbf{r}k,j}= \textbf{h}_{\textbf{r}k}^\text{H} y_k- \sum_{j' \in \xi_k \setminus j} \textbf{g}_{\textbf{r}k,j'} \hat{x}_{k,j'}^t
\end{align}    
    
    \item \emph{MMSE Filtering and LLR Computation }:
     To reduce the noise-plus-interference in (\ref{y_npi}), linear MMSE filters are used and using those, bit LLR from FN $k$ to VN $j$ is calculated. For the detailed MMSE-PIC algorithm, please refer \cite{scmapot_156, asic_mmsepic}. 
     
     %The MMSE-PIC matrices are given as
%\begin{align}
 %   \textbf{Z}_{\textbf{r}k}= \textbf{S}_{\textbf{r}k}^{-1}  \hat{\textbf{Y}}_{\textbf{r}k},\\
  %  \textbf{x}_{\textbf{r}k}= \textbf{S}_{\textbf{r}k}^{-1}  {\textbf{G}}_{\textbf{r}k},
%\end{align}
%where $\textbf{S}_{\textbf{r}k}= \textbf{G}_{\textbf{r}k} \Lambda_{\textbf{r}k} + \sigma^2 \textbf{I}_{d_f}$, $\Lambda_{\textbf{r}k}$ is a $d_f \times d_f$ diagonal matrix with $\lambda_{j,j}=\varepsilon_j$ and $\textbf{I}_{d_f}$ denotes the $d_f \times d_f$ identity matrix.

 %   \item \emph{LLR Computation}:
  %   Using the diagonal elements of $    \textbf{Z}_{\textbf{r}k}$ and $    \textbf{x}_{\textbf{r}k}$ denoted as ${Z}_{{r}k,j}$ and ${M}_{{r}k,j}$ for user $j$,  the bit LLR from resource node $k$ to user node $j$ is given as
%\begin{align}
 %    LU^{t,i}_{k \rightarrow j}= \frac{f_{i}(z_{rk,j})}{{M}_{{r}k,j}^{-1}- \varepsilon_{k,j}^t },
%\end{align}
%where $z_{rk,j}=\frac{{Z}_{{r}k,j}}{{M}_{{r}k,j}}$ and $f_{i}(z_{rk,j})= ( \min_{a \in \mathbb{X}_{b_j^i=0}} \vert z_{rk,j}-a\vert ^2 - \min_{a \in \mathbb{X}_{b_j^i=1}} \vert z_{rk,j}-a\vert ^2 )$.

\end{itemize}
%The mean and variance of codeword element $x_{k,j}$ %of $j$th user %($x_{k,j}$ is the codeword element from constellation set $\mathbb{X}$) 
%in the $t$-th SCMA iteration is given as
 In the last SCMA iteration, the \emph{a posteriori} bit LLR for the corresponding LDPC decoder is given as
\begin{align}
L_{j}^{P,i}= L_{j}^{A,i}+ \sum_{k' \in \zeta_j\setminus \{k\}} LU^{N_S,i}_{k' \rightarrow j}~,
\end{align}
% eqn 25 of scmapot_156
where $N_S$ is the maximum number of SCMA iterations. 
 To further improve the convergence rate, a modified user node update method is introduced, where user nodes pass the   \emph{a posteriori} bit LLR values to FNs instead of the extrinsic information. The MMSE-PIC based MP detection reduces the detection complexity by 93\% with a degradation of 0.63dB in SNR. 
 
\subsubsection{Polar coded SCMA}
\begin{figure}[t]
    \centering
    \subfigure[]{    \includegraphics[scale=0.40,trim=6cm 1cm 8cm 2cm,clip]{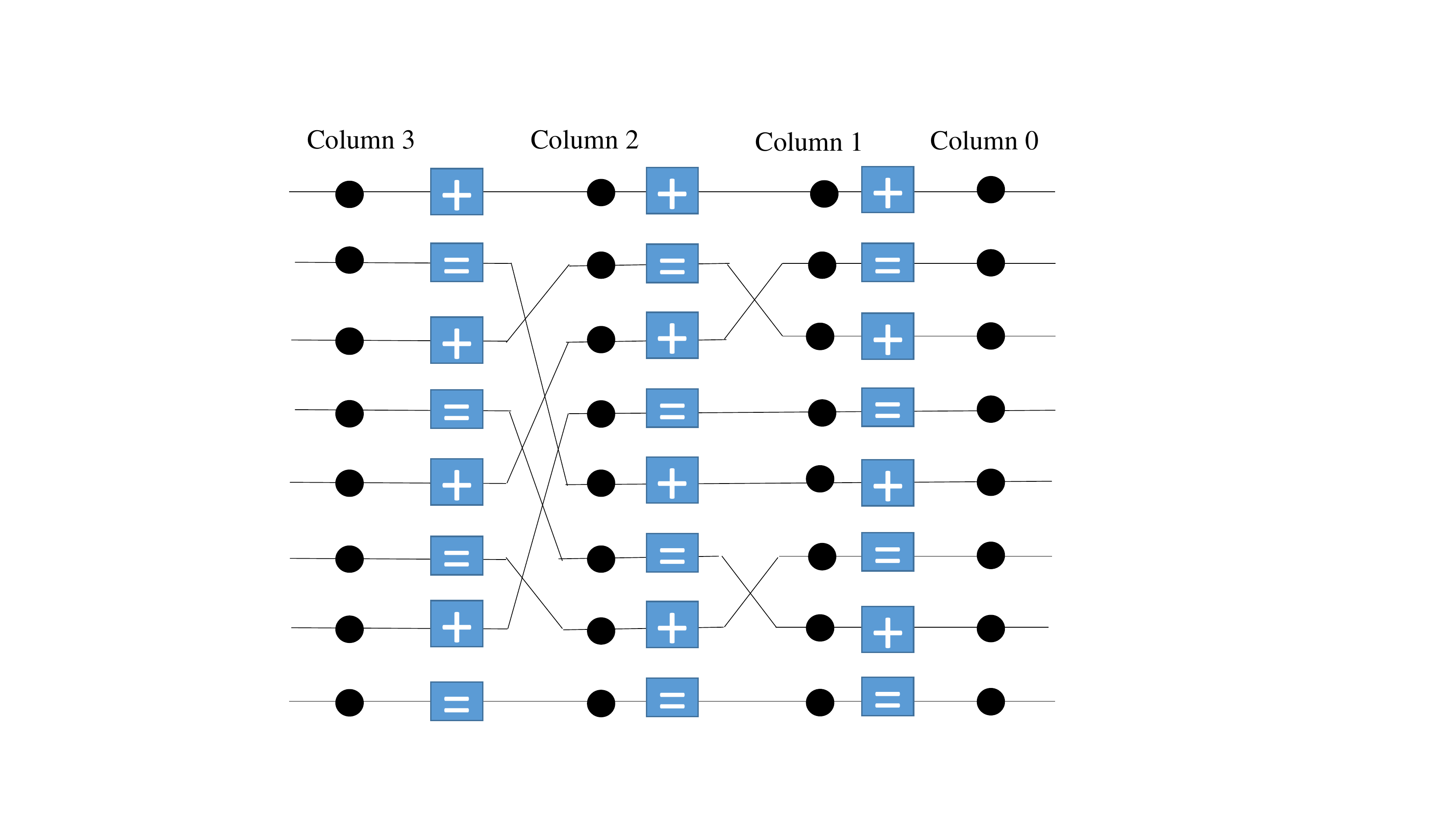}}
    %\label{fig:Mpa_ex_a}
    \subfigure[]{
    \includegraphics[scale=0.45,trim=3.5cm 6.5cm 17cm 2.5cm,clip]{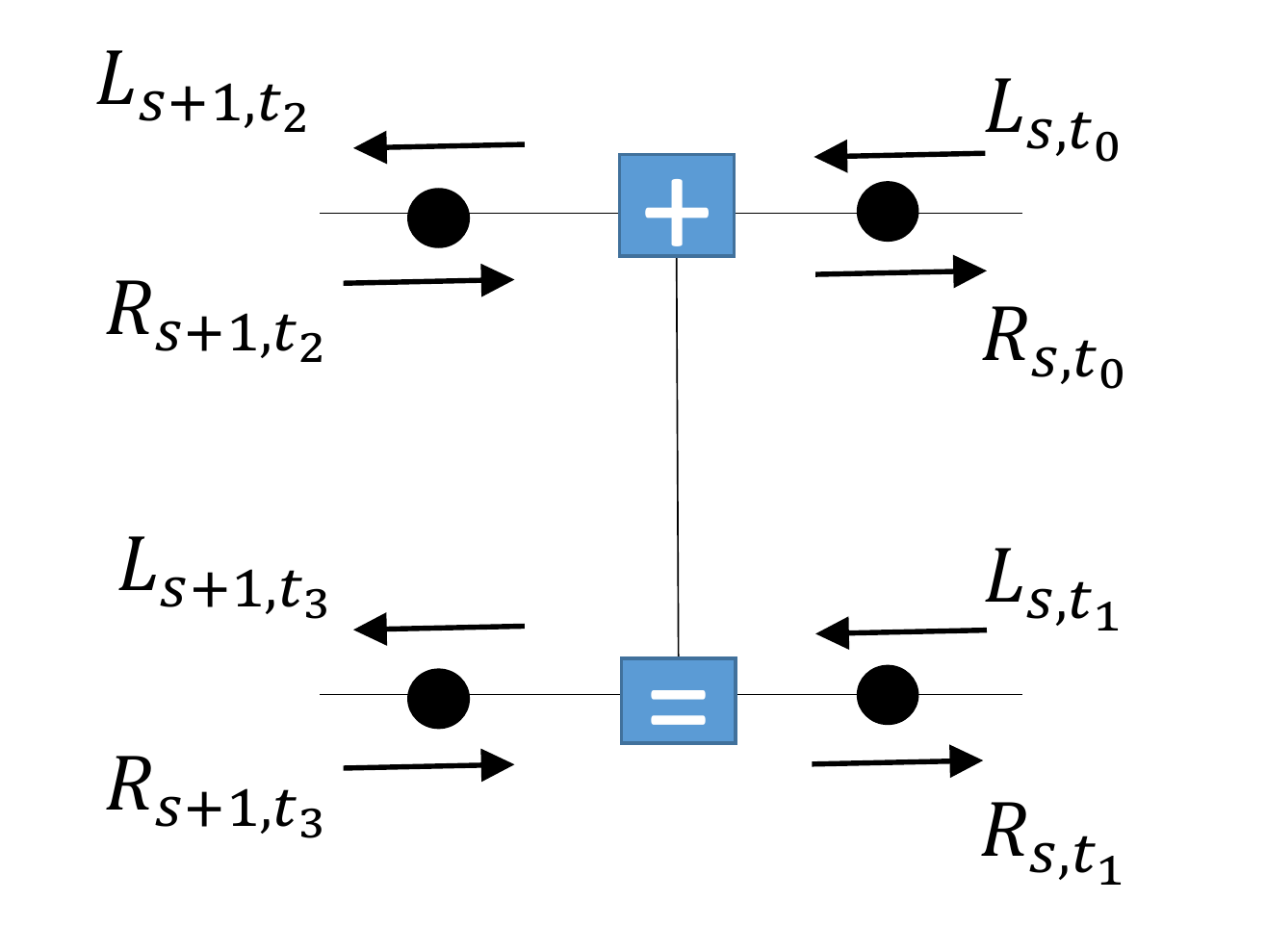}}
    %\label{fig:Mpa_ex_b}
    \caption{$(a)$ Factor graph for a polar code with $N$ = 8 $(b)$ Message passing
    on an unit graph.}
    \label{fig:polarfact_graph}
    \end{figure}

In practical scenarios, coded SCMA systems are employed to improve the quality of service (QoS). Recently, polar coded SCMA (PC-SCMA) has been studied, but the separate detection and decoding scheme does not give optimal performance. Thus, a joint iterative detection and decoding (JIDD) receiver was proposed for the uplink PC-SCMA system in \cite{scmapot_158_8}. SCMA detectors calculate
the extrinsic messages of the SCMA codewords of all users,
then the messages are demapped into a prior messages of each
user’s polar decoder. After polar decoders return the extrinsic
messages of each user’s polar codeword, they are remapped
to a prior messages of the SCMA codewords, which will be
passed to the SCMA detectors again.
This happens in one iteration of JIDD and  no inner iterations are needed for the SCMA detector and polar decoder. In the JIDD algorithm, the factor graphs of SCMA and polar encoder are combined to generate a  joint factor graph of the PC-SCMA system. The detection algorithm for SCMA is MPA, and of polar code is soft cancellation (SCAN) algorithm \cite{scmapot_158_14}. The JIDD algorithm is discussed below:
\begin{itemize}
    \item FNs update process:
    The FNs are updated and pass the information to neighboring VNs (as shown in Step 1 (a) of \textbf{Algorithm \ref{algo:on-grid2}}). Then, the bits extrinsic information in the form of LLR is calculated (as shown in Step 2 of \textbf{Algorithm \ref{algo:on-grid2}}).% This information is de-interleaved and passed as a prior information to the polar-decoder.
    %\begin{comment}
    %\begin{align}
     %   L_{a,polar} (\textbf{c}_j)= \Pi^{-1} (        L_{e,SCMA} (\textbf{b}_j))
    %\end{align}
    %\end{comment}
    \item Priori information update process: In this process, the output extrinsic information of the
FNs update process is passed to the polar decoder. Then,
a priori information of the VNs is calculated based on
the output extrinsic messages of the polar decoder.
    Each of the two adjacent columns of the polar code factor graph comprises $N/2$ unit factor graphs, where $N$ is the polar code length. The polar code factor graph with code length $N = 8$ is
shown in Fig. \ref{fig:polarfact_graph}$(a)$, which is composed of $\text{log}_2(N)$ + 1 = 4
columns, $N=8$ rows, and a unit factor graph is shown in Fig. \ref{fig:polarfact_graph}$(b)$. The factor graph has two associated left LLR ($L_{s,t}$) and right LLR messages ($R_{s,t}$)  at the node with the $s$th row index and $t$th column index. % The number of rows is $N$ and number of columns is $\text{log}_2(N)+1$ in the polar code factor graph. 
For initialization, the left message $L_{0,t}^j$ of the $j$th user is initialized with a prior information and the right messages are initialized as, $R_{s,t}=0 ~\text{for}~ t \in \mathcal{A}, R_{s,t}= \infty ~\text{for}~ t \in \mathcal{A}^c$, where $\mathcal{A}$ is the set of information bits and $\mathcal{A}^c$ is the set of frozen bits.  The messages computed to be passed on a unit factor graph are \cite{scmapot_158_14}:
    \begin{align}
        L_{s+1,t_2}^j=ms_f(R_{s+1,t_3}^j+L_{s,t_1}^j,L_{s,t_0}^j) \nonumber\\
                L_{s+1,t_3}^j=ms_f(R_{s+1,t_2}^j,L_{s,t_0}^j)+L_{s,t_1}^j\nonumber\\        R_{s,t_0}^j=ms_f(R_{s+1,t_2}^j,R_{s+1,t_3}^j+L_{s,t_1}^j)\nonumber\\        R_{s,t_1}^j=R_{s+1,t_3}^j+ms_f(R_{s+1,t_2}^j,L_{s,t_0}^j)
    \end{align}
    where $ms_f(a,b) \approx \text{sign}(a) \times \text{sign}(b) \times \text{min}(|a|,|b|)$ is called min-sum function.
    In Fig. \ref{fig:polarfact_graph}$(b)$, $ms_f (L_1,L_2)$ is denoted by $L_1 \boxplus L_2$ and is commonly referred to as boxplus operator and ``equal to" sign in the box denotes the addition operation. Thus, in each iteration, left and right messages are updated. For detailed polar SCAN algorithm, please refer  \cite{scmapot_158_14}. After one iteration of polar decoding, the output extrinsic information of the polar decoder is passed as a priori information to the SCMA detector. 
    \item VNs update process: In this, VNs update their information using a priori information   and pass it to the neighboring
FNs (similar to as shown in \textbf{Step 1} (b) of \textbf{Algorithm \ref{algo:on-grid2}}).

\end{itemize}
The above summarizes the JIDD algorithm for the PC-SCMA system. To further improve the performance, \cite{scmapot_158} used the EXIT chart to analyze the  convergence of the JIDD receiver. %Additionally, a weight factor is conceived and optimized for polar code construction for the PC-SCMA system. 
Under the 150\% overloading system, JIDD has 0.6dB performance loss over the Rayleigh fading channel with much less complexity. Further, a joint channel estimation and decoding (JCD) scheme 
has been proposed in \cite{scmapot_159} for the PC-SCMA system. This scheme did not require the instantaneous channel state information (CSI) over fading channels. In this, a sparse Bayesian learning based channel estimation algorithm was proposed to measure the initial CSI. Further, a soft-successive cancellation list decoding algorithm is proposed for polar decoding, which outputs LLR information to update the SCMA detector. Recently, an edge-cancellation-aided (EC-IDD) algorithm has been proposed for the PC-SCMA system, where Gaussian approximated
message passing (GA-MP) is for SCMA and  soft list decoding (SLD) for polar decoding\cite{pl_scma2021}.

%\fbox{\includegraphics[scale=0.5,trim=6cm 1cm 8cm 2cm,clip]{polar_factor_graph.pdf}}
%trim=left bottom right top
%\fbox{\includegraphics[scale=0.5,trim=3.5cm 6.5cm 17cm 2.5cm,clip]{unit_polar_factor_graph.pdf}}

\subsection{SIC-MPA}
In \cite{scmapot_112}, two successive interference cancellation (SIC) based receivers, namely SIC and joint SIC-MPA receivers, were proposed for SCMA systems. The complex symbols are divided into real and imaginary parts, and the system model is rewritten to make a simpler receiver design. SIC receiver consists of subtraction and deduction operations and thus, has lower complexity than MPA. However, the SIC receiver suffers from performance loss. In the joint SIC-MPA receiver, the output of SIC is sent to MPA to calculate LLR. Since SIC provides an initial value to MPA, this receiver converges faster, and provides better performance improvement of 0.5-1 dB with less complexity. However, the main trade-off is that specific CBs are required for this receiver, and the proposed decoding scheme cannot be used as it is in the generally used RF CBs. %(such as described in \cite{scmapot_91}). 

\subsection{Other Methods for SCMA Detection}
%\section{Expectation Propagation Based SCMA Detection}
The classical decoding algorithm for SCMA is MPA which provides ML-like performance with complexity $M^{d_f}$. In literature, some schemes different from MPA have been proposed  for SCMA detection.  

\subsubsection{Expectation Propagation}
The expectation propagation algorithm (EPA) is an approximate Bayesian inference method. EPA can be used for estimating  posterior distributions
with simple distributions through distribution projection\cite{scmapot_130_10,scmapot_130_11}. A low complexity iterative receiver based on EPA  was proposed for uplink SCMA system in \cite{scmapot_130,scmapot_133}. With EPA, the decoding complexity reduces from exponential to linear, i.e., it is linearly proportional to the CB size $M$.

The projection of a  distribution
$p$ onto some distribution set $\Phi$ is given as
\begin{align*}
    \text{Proj}_{\Phi}(p)= \mathop{\arg \min}_{ q \in \Phi} \text{KL}(p ||q),
\end{align*}
where $\text{KL}(p ||q)$ denotes the Kullback-Leibler (KL) divergence.
In EPA, the messages passing
between VNs and FNs can be considered as continuous RVs. Next, the messages in the \textbf{Step 1} of \textbf{Algorithm 1} can be simplified to scalar complex Gaussian
distributions, and this step can be replaced  by updating the mean and variance parameter. After calculating the mean and variance of the messages passed between the nodes for multiple iterations, the LLR can be calculated.
A new initialization method  to accelerate the
convergence as well as reducing the complexity of the algorithm was proposed in \cite{scmapot_133}. \color{black}In \cite{epa_ldpc_scma}, a  receiver design by integrating EPA detector with JDD was proposed for an uplink LDPC-SCMA system.%Simulation results have shown that EPA receiver achieves nearly the same block error rate (BLER) performance as the standard MPA receiver with much less complexity.
\color{black}

\subsubsection{Deep Learning Based SCMA Detectors}
Deep learning is a subclass of machine learning which has contributed to many aspects of wireless communication systems, such as channel estimation, signal detection, resource allocation, network security, etc \cite{scmapot_160,scmapot_160_11}. Recently, deep learning was applied for SCMA detection in which Deep Neural Networks (DNN) were  trained to provide non-iterative decoding with much less complexity \cite{scmapot_169, scmapot_170, scmapot_171, scmapot_172,scmapot_174,dnnmpa_2021}. %In \cite{scmapot_174}, denoising autoencoders (DAE) are use to jointly design the SCMA CB and the detector. 
%Let $J$ users be transmitting data over $K$ REs, and the  complex data is converted into real and imaginary parts.  Thus, 
In \cite{scmapot_174}, DNN based SCMA decoder has $2K$ nodes at the input layer and $\text{log}_2(M) J$ nodes at the output layer. The aim of training the DNN based SCMA decoder is to optimize the weights  and biases, such that the cross entropy between the original transmitted bits and the output of the DNN decoder is minimum. %This can be expressed as
%\begin{align}
 %   \min_{\textbf{W}_d,\textbf{B}_d} \biggl( - \sum_{i'=1}^{\text{log}_2(M)J } b'_{i'} \text{log} \big(\pi_i \big[d \big( ( y_1,\cdots,y_{2K})^T;\textbf{W}_d,\textbf{B}_d \big) \big] \big)  \biggr)
%\end{align}
%where \hl{$\textbf{b'}=[b'_1,\cdots,b'_{\text{log}_2(M)J}]$ denote the transmitted bits}, the received signal $\textbf{y'}=[y'_1,\cdots,y'_{2K}]$, and $d(.; \textbf{W}_d,\textbf{b}_d)$ denotes the DNN decoding process. 
The received signal is of dimension $2K$ since each  received signal at $k$th RE, i.e., $y_k$ is divided into real and imaginary parts.   % For simulation, gradient descent training optimizer ADAM is adopted.  
The DNN based SCMA decoder performance is slightly poor than the standard MPA decoding algorithm. So, \cite{scmapot_174}  designed denoising autoencoders (DAEs) for jointly designing SCMA CB and  decoder. Autoencoder based SCMA (AE-SCMA) outperforms the standard MPA scheme, and its performance is close to dense code multiple access (DCMA) for low SNR regions\cite{dcma_zi}. \color{black} Table \ref{table:one} summarizes the approach, the advantages and disadvantages of the SCMA detection
algorithms.% comparing their pros and cons.

\begin{table*}[htbp]
\caption{Review of SCMA detector algorithms.}
\centering
\begin{tabular}{|c|p{0.1\linewidth}|p{0.22\linewidth}|p{0.2\linewidth}|p{0.2\linewidth}|}%{|M{1.25cm}|M{2 cm}|M{4 cm}|M{3 cm}|M{3 cm}|} 
\hline \rule{0pt}{3ex} \textbf{Reference} & \textbf{Approach} & \textbf{Description} & \textbf{Advantages} & \textbf{Disadvantages}  \\ \hline \hline
 \rule{0pt}{3ex}\cite{lds} & Standard MPA in NOMA & Proposed MPA for low-density signature CDMA (LDS-CDMA) scheme. & LDS-CDMA
improves system performance by exploiting
LDS sequences in CDMA & LDS-CDMA lacks the ``shaping gain" as compared to SCMA. 
 \\ \hline
\rule{0pt}{3ex} \cite{scmapot_121_35}, \cite{scmapot_121}& PM-MPA & The codewords of a given number of users is detected after a number of iterations. In the remaining iterations, the codewords of the remaining users is determined. &   Number of collision users on each RE can be reduced. &  
\begin{minipage}[t]{0.4\textwidth}
    \begin{itemize}
    \item {Slight  BER\\
performance \\degradation.}
    \item  {CSI needs to\\ be known.}% MUSA PAper
    
    %\item  {Spreading sequences\\ have low \\cross-correlation.}
    \end{itemize}
  \end{minipage}\\ \hline \rule{0pt}{3ex}\cite{scmapot_112} & SIC-MPA & SIC and MPA are combined. & MPA
iteration times are decreased. & Specific CBs are required.\\ \hline \rule{0pt}{3ex}
  \cite{scmapot_108},\cite{scmapot_109} & Max-log-MPA &  Approximating the logarithm of the sum of exponential operation to a maximum operation using Jacobian logarithm. & 
Complexity arises from only addition operations $\mathcal{O}(2K M^{d_f} d_f)$. & Slight  BER
 degradation and CSI needs to be known. \\ \hline \rule{0pt}{3ex}
\cite{scmapot_140}, \cite{scmapot_141} & DMPA &
Considered the variable nodes as random variables on complex field and then,
applied discretization and FFT to update the messages. & Detection complexity
per FN reduces to $ O(d_f^3
\text{ln}(d_f))$. & Discretization  causes approximation error and CSI needs to be known. \\ \hline
% \cite{scmapot_144},   
\rule{0pt}{3ex}
\cite{scmapot_151} & Sphere Decoded MPA (SD-MPA) & Applied Log-MPA algorithm while updating the FNs within a restricted region only. & Applicable for regular and irregular constellation topology.& If the unitary rotation angle of the hypercube increases than a certain value, BER degrades. Also, CSI needs to be known. \\ \hline
 \rule{0pt}{3ex} \cite{scmapot_111} & Scheduled MPA & Simplified  MPA from two aspects. Firstly, a lookup table (LUT) was designed, and secondly, the FNs and VNs are jointly updated in SS-MPA. In MS-MPA, number of SS-MPA detectors work in parallel with different update orders. & Better error performance and
higher convergence speeds. & High complexity and CSI needs to be known. \\ \hline
 \rule{0pt}{3ex} \cite{scmapot_122},\cite{scmapot_123} & Extended max-log MPA & Dynamic trellis based MPA is proposed such as the truncation is decreased as the iterations progresses. & Low complexity $O(m_c^{d_f})$. & Slight BER degradation and CSI needs to be known.
 \\ \hline 
%\end{tabular}
%\label{table:one}
%\\  \hspace{12cm}{\footnotesize(To be continued)}
%\end{table*}

%\begin{table}[htbp]
%\caption{Complexity analysis of MPA and Max-Log-MPA considering shot noise.}
%\centering
%\begin{tabular}{|M{1.25cm}|M{2 cm}|M{4 cm}|M{3 cm}|M{3 cm}|} \hline  \textbf{Reference} & \textbf{Approach} & \textbf{Description} & \textbf{Advantages} & \textbf{Disadvantages}  \\ \hline \hline
\rule{0pt}{3ex} \cite{scmapot_158_8,scmapot_158, scmapot_159} & JIDD for Polar SCMA & A joint iterative detection and decoding was proposed where  MPA and  soft cancellation (SCAN) algorithm are used for SCMA and polar decoding, respectively. & Only needs outer iterations and no inner iterations of constituent SCMA detector and polar decoder are needed. & Slight BER degradation. \\ \hline
 \rule{0pt}{3ex} \cite{scmapot_156} & MMSE-PIC for LDPC-SCMA & Optimized
the interface between the SCMA detector and the
LDPC decoder resulting in a feasible  hardware solution. & Low complexity and implemented using 40nm CMOS technology. & Not investigated for high-order modulations. \\ \hline
 \rule{0pt}{3ex} \cite{scmapot_130} & EPA & An approximate Bayesian inference method, where message passing is reduced to mean and variance calculation of approximate Gaussian distributions.& Complexity  scales linearly with $M$ and $d_f$.& Not investigated for large scale systems and CSI needs to be known. \\ \hline
  \rule{0pt}{3ex} \cite{scmapot_169}, \cite{scmapot_174} & DL-SCMA & Autoencoders are  designed to generate the CBs and the detectors for the AWGN channel.& Automatically
learn to construct optimal CBs and the corresponding decoder.& DL-SCMA performance degrades with predefined CBs.
 
 \\ \hline

\end{tabular}
\label{table:one}
\end{table*}

 %DCMA is a generalized version of SCMA in which most or all sequence entries are non-zero in the factor graph matrix \cite{dcma_zi}.

\subsection{MIMO SCMA Detection}
The spectral efficiency of an SCMA system can be further increased by integrating it with another key technique of  the 5G wireless
system, i.e., multiple-input
multiple-output (MIMO). However, the detection complexity of standard MPA for the MIMO-SCMA increases exponentially
with the number of antennas and accessed users. Some
research works proposed MIMO-SCMA receivers to reduce the detection complexity \cite{scmapot_133,scmapot_175,reviewer2_ref2,reviewer2_ref1,scmapot_177,scmapot_178,scmapot_129,scmapot_131}. 

In \cite{scmapot_175},   MPA detection 
based on QR decomposition was proposed for the uplink SCMA system
with multiple  antennas at the BS. Specifically, the number of FNs depends only
on  $d_f$, which reduces the detection complexity.  In \cite{reviewer2_ref2}, Gaussian approximation
based belief-propagation (GA-BP) detection with linear prefiltering was proposed, along with the diversity analysis. This scheme has low  computational complexity but suffers from performance degradation\cite{disadvantage_BP}. To improve the convergence
of GA-BP detection, \cite{reviewer2_ref1} proposed a CB design that also improves the BER performance. In \cite{scmapot_177}, a partially active MPA scheme for the MIMO-SCMA was proposed. A sliding window was introduced to find the active  users during
each iteration while others remain silent. Thus, the detection complexity  efficiently reduces
while its performance  degrades slightly. This research work was extended in \cite{scmapot_129}, where two Gaussian-approximated MPA (GA-MPA)
schemes are proposed for detection. To be specific, the output from the decoders were used to estimate the mean and variance of the Gaussian messages in the detector, and a damping factor was introduced  to improve the performance loss. The detection complexity is linearly related to the number of
antennas and $d_f$. In \cite{scmapot_178},  a resource-selection based MPA (RSB-MPA) was proposed for uplink SCMA system with multiple antennas at BS. The RSB-MPA is applied at well-conditioned
resources and Maximum Likelihood algorithm was applied at resources with bad conditions. This resulted in a trade-off between the BER performance and complexity.
In  \cite{scmapot_131},  a stretched factor graph based 
hybrid BP-EP
receiver was proposed for a downlink MIMO-SCMA system
over frequency selective fading channels. 
However, this scheme
can not be applied to users with multiple antennas. In \cite{scmapot_133}, 
a sparse-channel based EPA (SC-EPA) was proposed for uplink MIMO-SCMA system. The complexity of the 
SC-EPA is independent of the number of  antennas at the BS
and is linearly dependent on   the constellation size. Also, it is more robust to channel estimation error.
\\
Below are a few of the  challenges
of SCMA  detection algorithms, along with opportunities to further improve the SCMA systems:
%and future research trends addressing these challenges
\begin{itemize}
\item \emph{Complexity}: SCMA    may outperform  other NOMA schemes, but the standard MPA 
decoding could be unaffordable  in terms of complexity \cite{scmapot_23}.  More research works are needed  to  further reduce the decoding complexity especially for large-sized SCMA systems.     \item \emph{Channel Estimation:} Most of the aforementioned
detection methods assume that the CSI is perfectly  known; however, this assumption is not realistic.  Few works studied a joint channel
estimation and data decoding methods \cite{scmapot_136,scmapot_159}. The
impact of CSI estimation errors to different detection methods remains largely unknown. 

\item \emph{Synchronization:} The SCMA detectors discussed above assume a
perfect synchronization at the receiver. In practice, there may be  performance degradation in presence of asynchronous transmission  \cite{scmapot_135}. %For instance, a delayed signal from a user will disturb the message exchange in MPA based detectors. 
Thus, more
robust detectors which can cope  with synchronization
errors are expected. 
\item \emph{Adaptive SCMA:} In most of the current SCMA systems, all the users are uniformly
served, which may not be the optimal strategy.  Practically, it is desirable that SCMA codebooks and detectors  can be 
adapted to different users requirements. Some existing works on adaptive SCMA
are proposed in \cite{scmapot_69,scmapot_70}. 

In addition to the above points, the SCMA detectors should also be designed for emerging wireless technologies, such as  beamforming, massive number
of antennas in MIMO, multi-user MIMO,
 reconfigurable intelligent surface (RIS), etc \cite{mumimo_reviewer1}.
 
\end{itemize}

\color{black}
\section{New Applications and Future Directions}
%With the target of achieving higher transmission rates, low latency,  and better user experience, future wireless applications include massive machine type communications (mMTC), ultra dense networks,  and ultra reliable low latency communications (URLLC) services. SCMA is a code domain NOMA scheme which can greatly improve the network capacity with different overloading factors. 
This section presents a few promising directions for future SCMA research:%, which may be of interest to enhance the system capability in B5G and 6G.
\begin{itemize}
    \item \emph{Grant Free SCMA:}
    With the deployment of massive number of wireless devices in  various vertical industries (such as factory automation, autonomous driving, smart health care, etc.), massive machine-type communication (mMTC) has been attracting tremendous research attention over the past years.
With mMTC, one aims to achieve both massive connectivity and low latency. Legacy communication systems are highly inefficient due to the significant  signaling overhead and long handshaking latency imposed by the grant based scheduling\cite{grantfree_1_1}.
Grant-free access has been extensively studied to make a difference where the active users contend the
shared resources by directly transmitting to the BS. For grant-free SCMA, the following two problems are worth a more profound investigation:  active user identification and collision resolution.

In most of the works on SCMA, it is assumed that the channel state information (CSI) is known, and all the users  transmit data simultaneously. However, this requires the BS to identify the active users before decoding. An efficient way for active user detection and channel estimation is to formulate the
two problems jointly as a sparse signal recovery problem
and then attack it by the compressed sensing theory.

A receiver that performs joint channel estimation,
data decoding, and the detection of active users based
on EP message passing was proposed in \cite{towards5g,grantfree_4,grantfree_5}.  Also,  a blind detection and compressive sensing based receiver was  proposed in \cite{grantfree_3} for performing the joint activity and data detection. Grant-free SCMA allows users to transmit data in the preconfigured resources, which is called contention transmission unit (CTU), i.e., a combination
of time, frequency, CB, and pilot. In \cite{grantfree_2}, an acknowledgment  feedback based user-to-CTU mapping was proposed for grant-free SCMA to decrease the collision probability.

\item \emph{SCMA for High Mobility Communications:} With the rapid evolution in autonomous driving, the prevalent vision is that connected
autonomous vehicles will be seen everywhere in the next decade \cite{6gpaper_15}.
Challenges may arise when a vehicle moves at a very  high
speed (e.g., 500 km/h or higher) and exchanges information
with its surrounding vehicle or infrastructure using SCMA. In this case, innovative solutions are needed to maintain an excellent reception quality. One direction is to study the integration of SCMA
with orthogonal-time-frequency-space (OTFS). OTFS is a 2D-modulation technique with the allocation of symbols on the delay-Doppler domain that shows superior performance
in high mobility scenarios. It is expected that the OTFS-SCMA
system will provide massive connectivity in the
high-speed vehicle-to-everything (V2X) scenario. In \cite{otfs_scma_1}, a multi-user OTFS system based on SCMA was proposed, and receivers were designed for both uplink and downlink systems. 

\item \emph{SCMA for Visible Light Communications:} 
Demands for data-intensive applications in future  wireless applications have driven visible light communications (VLC) as a promising  candidate to
provide high-speed and secure
indoor/outdoor wireless access across unlicensed
spectrum \cite{scmapot_220,scmavlc_demons_1,scmavlc_demons_2}. SCMA has been employed to enhance the bandwidth efficiency in VLC  systems. The VLC systems have the requirement of  real and positive signal transmission.  In \cite{vimal_12,vimal_13,vimalpaper},  radio-frequency (RF)  based multi-dimensional CBs have been used and made suitable changes  so that they can be applied to the VLC system. The SCMA CBs have been designed specifically for VLC systems in \cite{vimal_8,rmedpaper,dropaper, saumya_cb_design}.

\item \emph{SCMA with Reconfigurable Intelligent Surface:}
RISs  have received significant attention from both academia and industry  as it can  smartly reconfigure the wireless propagation
environment\cite{scmapot_228}. RIS improves the
coverage performance of the wireless system by reflecting the incident
signals after adjusting their phases and/or amplitudes in the desired direction \cite{risscma_2_2}. The integration of  RIS and SCMA was investigated in \cite{risscma_decoder,risscma_phase} where RIS phase shifts are optimized, and a low-complexity decoder was designed to improve the system performance.   

\end{itemize}

\section{Conclusion}
This paper has provided a systematic self-contained tutorial for SCMA, a disruptive code-domain NOMA scheme for enabling massive connectivity.   
SCMA uses carefully designed sparse CBs for significant constellation shaping gain and anti-interference capability  compared to previously proposed code-domain NOMA techniques. Compared to the complex MAP decoding, iterative MPA gives rise to efficient MPA decoding with significantly reduced complexity. Thus, SCMA represents a promising solution for  providing better quality of service to users (with the overloading factor greater than one), low latency and high spectral efficiency. It is anticipated that this paper will provide a quick and comprehensive understanding of SCMA to motivate more researchers/engineers toward this research topic.

\end{document}